\documentclass[a4paper,fleqn,usenatbib]{mnras}

\usepackage[T1]{fontenc}
\usepackage{ae,aecompl}

\usepackage{graphicx}	% Including figure files
\usepackage{amsmath}	% Advanced maths commands
\usepackage{amssymb}	% Extra maths symbols
\usepackage{pdflscape}
\usepackage{longtable}
\usepackage{caption}
\usepackage[T1]{fontenc}
\usepackage{placeins}

\newcommand{\sfg}{\,S$^4$G}
\newcommand{\eblg}{\,$\epsilon Blg$ }

\defcitealias{2015ApJS..219....4S}{S15}

\title[Non-parametric decompositions of disk galaxies]{Non-parametric decompositions of disk galaxies in \sfg{}   using \textsc{DiskFit}}

\author[Lewis \& Spekkens]{
C. Lewis,$^{1}$
K. Spekkens,$^{1,2}$
\\
$^{1}$Department of Physics, Engineering Physics and Astronomy, Queen's University, Kingston, ON K7L 3N6, Canada\\
$^{2}$Department of Physics and Space Sciences, Royal Military College of Canada, PO Box 17000, Station Forces, Kingston, ON K7K 7B4, Canada\\}

\date{Accepted XXX. Received YYY; in original form ZZZ} 
\pubyear{2018}

\begin{document}
\label{firstpage}
\pagerange{\pageref{firstpage}--\pageref{lastpage}}
\maketitle
\begin{abstract}
	We present photometric models of 532 disk galaxies in 3.6$\mu$m images from the Spitzer Survey of Stellar Structure in Galaxies (\sfg{}) using the non-parametric \textsc{DiskFit} algorithm. We first test \textsc{DiskFit}'s performance on 400 synthetic \sfg{}-like galaxy images. \textsc{DiskFit} is unreliable in the bulge region, but accurately disentangles exponential disks from Ferrers bars farther out as long as their position angles differ by more than $5^\circ$. We then proceed to model the \sfg{} galaxies, successfully fitting 489 of them using an automated approach for initializing \textsc{DiskFit}, optimizing the model and deriving uncertainties using a bootstrap-resampling technique. The resulting component geometries and surface brightness profiles are compared to those derived by Salo et al using the parametric model \textsc{galfit}. We find generally good agreement between the models, but discrepancies between best-fitting values for individual systems are often significant: the choice of algorithm clearly impacts the inferred disk and bar structure. In particular, we find that \textsc{DiskFit} typically assigns more light to the bar and less light to the disk relative to the Ferrers and exponential profiles derived using \textsc{galfit} in the bar region. Given \textsc{DiskFit}'s reliability at disentangling these components in our synthetic images, we conclude that the surface brightness distributions of barred \sfg{} galaxies are not well-represented by these functional forms. The results presented here underscore the importance of validating photometric decomposition algorithms before applying them to real data and the utility of \textsc{DiskFit}'s non-parametric approach at measuring the structure of disks and bars in nearby galaxies.
\end{abstract}


%%%%%%%%%%%%%%%%%%%%%%%%%%%%%%%%%%%%%%%%%%%%%%%%%%

%%%%%%%%%%%%%%%%% BODY OF PAPER %%%%%%%%%%%%%%%%%%

\section{Introduction}
Secular processes dominate the evolution of galaxies at low redshifts, with galactic bars being among the main mechanisms driving evolution \citep[e.g.][]{2004ARA&A..42..603K,2013seg..book....1K,2013seg..book..305A,2014RvMP...86....1S}. Bars are long-lived features, initially formed as elongated gravitational instabilities in the plane of the galaxy disk \citep[e.g.][]{1971ApJ...168..343H,1981A&A....99..362S,1987gady.book.....B}. As the instability grows and the bar gets stronger it will increase in length and trap more disk stars. This takes angular momentum from the bar, making it thinner and decreasing its rotational pattern speed \citep{2003MNRAS.341.1179A}. Absorption of angular momentum by the halo also strengthens the bar \citep{2002ApJ...569L..83A,2003MNRAS.341.1179A}. Additionally, bars dissipate angular momentum via gas in the disk, channelling it toward the centre of the galaxy and triggering central star formation \citep{1997MNRAS.287...57E,1992MNRAS.259..328A,2013pss5.book..923S,Tonini2016}. The physical properties of bars are therefore important indicators of the evolutionary states of their host galaxies \citep{2017MNRAS.470L.122P,2018MNRAS.473.4731K,2011MNRAS.415.3627H,2011MNRAS.411.2026M,2013MNRAS.429.1949A}.

Over the years, a variety of techniques have been adopted to extract bar properties from photometric images of nearby galaxies. Initial attempts focused on applying parametric functions to one-dimensional surface brightness profiles derived by fitting isophotal ellipses \citep[e.g.][]{1977ApJ...214..359K,1977ApJ...217..406K,1977ApJ...218..333K,1985ApJS...59..115K,1979ApJ...234..435B}, and changes in the ellipticity and position angle of the ellipses themselves can also be used to characterize bars \citep{1995A&AS..111..115W,2005MNRAS.364..283E}. However, tests show that the latter approach leads to systematic biases in the resulting bar lengths \citep{2008MNRAS.384..420G,2009A&A...495..491A}, while significant parameter degeneracies can emerge from the former approach, particularly near the galaxy centre \citep[e.g.][]{2003ApJ...582..689M,1977ApJ...217..406K}.

More recently, applying two-dimensional models directly to the imaging data has become the norm for characterizing bars. Several algorithms (e.g. \textsc{budda}, \citealt{2004ApJS..153..411D}; \textsc{gim2d}, \citealt{1998ApJ...507..585M,2002ApJS..142....1S}; \textsc{imfit}, \citealt{2015ApJ...799..226E}; \textsc{galfit}/\textsc{galfitm}, \citealt{2002AJ....124..266P,2010AJ....139.2097P,2013MNRAS.435..623V}) decompose galaxy images into parametrized disks, bars, and bulges, and more complex structural components such as arms and rings can also be incorporated \citep{2010AJ....139.2097P,2017ApJS..230...14M,2017MNRAS.466..355M}. The reliability and uniqueness of these models is hard to infer, however, since many algorithms don't return uncertainties on best-fitting values. Bar parameters are thus often presented without uncertainties \citep{2008ASPC..393..279W,2015ApJS..219....4S,2015ApJ...808...90D}. Although modelling large galaxy samples is an effective means of minimizing the statistical uncertainties on measured values \citep{2005AJ....129.2562B}, detailed survey- and algorithm-specific tests are required to determine whether or not systematic biases are important \citep{1987AJ.....93...60S,1995ApJ...448..563B,1999AJ....117.1219W,2003ApJ...582..689M,2008MNRAS.384..420G,2017MNRAS.469.3541P}.

\textsc{DiskFit} is an algorithm for modelling two-dimensional velocity maps or images of nearby disk galaxies \citep{2007ApJ...664..204S,2015arXiv150907120S,2012MNRAS.427.2523K}. It was designed to measure the kinematic and photometric properties of asymmetries in these systems and to generate robust uncertainties of the best-fitting values using a bootstrap-resampling technique \citep{2010MNRAS.404.1733S}. In the uncertainty estimation, the residuals from a best-fitting model are randomly scrambled and added to the model to create each bootstrap realization. The standard deviation of the best fitting model values from the ensemble of realizations is then adopted as the uncertainty \citep{2007ApJ...664..204S}. The photometric side of the code differs from many other approaches in that it models the surface brightness distribution of the disk and bar -- each assumed to have a fixed centre, position angle and ellipticity -- non-parametrically.  \textsc{DiskFit} is therefore particularly useful for characterizing the properties of barred galaxies whose bar and disk light distributions may differ from the standard Ferrers and exponential forms, respectively, as suggested both theoretically and observationally in recent studies \citep{2008MNRAS.384..420G,2016MNRAS.462.3430K,2017MNRAS.469.4414W}. 

The Spitzer Survey of Stellar Structure in Galaxies (\sfg{}, \citealt{2010PASP..122.1397S}) produced 3.6$\mu$m and 4.5$\mu$m images for 2331 nearby galaxies in order to map their stellar mass distributions at wavelengths largely immune to dust contamination \citep{2013ApJ...771...59M,2015ApJS..219....5Q,2015ApJS..219....3M}. The \sfg{} sample is therefore an ideal testbed for comparing the performance of different photometric decomposition models at recovering disk and bar properties.  \citealt{2015ApJS..219....4S} (hereafter \citetalias{2015ApJS..219....4S}) present multi-component ``human-supervised'' models of \sfg{} barred and unbarred galaxies using \textsc{galfit} that have been extensively used to constrain their underlying structure and evolution \citep{2015ApJS..219....3M,2015MNRAS.451.1004E,2016MNRAS.455.2644S,2017MNRAS.471.1070B}. \citetalias{2015ApJS..219....4S} discuss the impact of changing the PSF, sky subtraction, weight smoothing, and model components on the best-fitting \textsc{galfit} models, and also compare to models in the literature for a subset of the sample. However, \citetalias{2015ApJS..219....4S} do not report uncertainties on their final parameters or test \textsc{galfit}'s performance on \sfg{}-like imaging, and the reliability of their decompositions is therefore not quantified.  Since \textsc{DiskFit}'s non-parametric approach to characterizing disk and bar components differs from \textsc{galfit}'s parametric one, it is an ideal algorithm to compare to the \citetalias{2015ApJS..219....4S} results to gain a better understanding of the underlying structure of nearby disks and bars and the uncertainties associated with extracting this structure from imaging data (e.g. \citealt{2010AJ....139.2097P}). 

This paper presents non-parametric \textsc{DiskFit} decompositions of \sfg{} disk galaxies for direct comparison with \citetalias{2015ApJS..219....4S}'s parametric \textsc{galfit} models. Our goals are to quantitatively test \textsc{DiskFit}'s performance on \sfg{}-like imaging, to determine the conditions under which \textsc{DiskFit} and \textsc{galfit} models of \sfg{} galaxies differ, and to explore the resulting implications for the structure of barred galaxies. \S\ref{section2} describes the \sfg{} subsample adopted for our comparison. \S\ref{section3} describes the synthetic galaxy images that we used to test \textsc{DiskFit} on \sfg{}-like imaging and the corresponding modelling results. \S\ref{section4} presents our \textsc{DiskFit} models of the \sfg{} subsample itself, and a comparison between the best-fitting properties of these models to the corresponding \citetalias{2015ApJS..219....4S} \textsc{galfit} results. We discuss the implications of that comparison for the structure of nearby barred galaxies in \S\ref{section5}.

\section{Sample Selection}\label{section2}
\begin{figure}
 \includegraphics[width=\columnwidth]{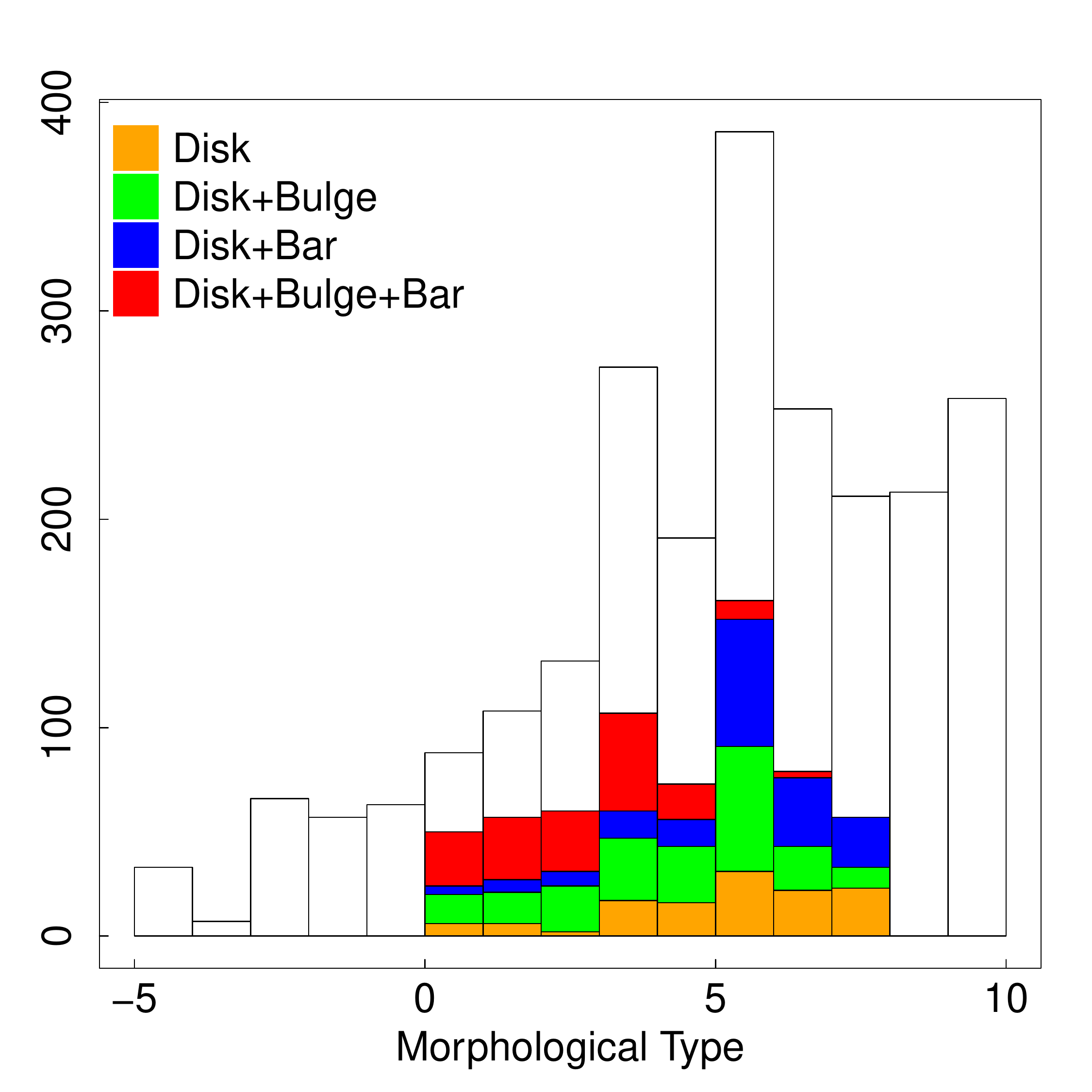}
	\caption{Distribution of morphological types included in the \sfg{}   subsample modelled with \textsc{DiskFit} (coloured bars), compared to that of all \sfg{}   galaxies (open bars). The coloured bars show subsample galaxies with different structural components as determined by the \textsc{galfit} models of \citetalias{2015ApJS..219....4S}: the orange histogram represents galaxies that have a only a disk, the green histogram represents galaxies that have a disk and a bulge, the blue histogram represents galaxies that have a disk and a bar, and the red histogram represents galaxies that have a disk, a bar, and a bulge. \label{Figure1} }
 \end{figure}

There are 2331 galaxies in \sfg{}   \citep{2010PASP..122.1397S}, encompassing all morphological types. Very early and late type galaxies tend not to have distinct disks, and are not well represented by \textsc{DiskFit}'s models. We therefore include only rotationally supported spirals (Hubble types 0 through 7) in our subsample, as listed in the \sfg{}   catalogue \citep{2010PASP..122.1397S}. Since our focus is the characterization of bars in nearby, well-resolved disk galaxies, we exclude high inclination systems ($\epsilon Disk$ > 0.7) and low-quality images (quality flag $q$ $<$ 4 from \citetalias{2015ApJS..219....4S}), and require that the scale length of the disk measured by \citetalias{2015ApJS..219....4S} be spanned by at least 15 pixels (1 pixel = 0.75 arcseconds; \citealt{2010PASP..122.1397S}). The \sfg{}   subsample that meets these selection criteria contains 570 galaxies, some of which are listed in Table \ref{table1} (the full table is shown in digital appendix Table A1). We fit the same 3.6$\mu$m \sfg{} images as modelled by \citetalias{2015ApJS..219....4S}. Figure \ref{Figure1} shows the distribution of morphological types included in the subsample, and Figure  \ref{Figure2} shows the distribution of structural parameters in that subsample inferred from the \textsc{galfit} models of \citetalias{2015ApJS..219....4S}.
%\onecolumn
\begin{table*}
\centering
\begin{tabular}{ccccccccc}
\hline
Name            & Model & Disk PA [$^\circ$]        & $\epsilon$Disk        & Bar PA [$^\circ$] & $\epsilon$Bar & Disk \%        & Bar \%       & flags\\
(1) & (2) & (3) & (4) & (5) & (6) & (7) & (8) & (9) \\
\hline
ESO012-010      & dbar  & $151(1)$ &  $0.62(0.02)$ &  $24(3)$ &  $0.44(0.04)$ &  $88(1)$ &  $12(1)$ &  \\
ESO013-016      & dbar  & N/A   & N/A   & N/A   & N/A   & N/A   & N/A   & a \\
ESO026-001      & dbar  & $45(14)$ &  $0.25(0.07)$ &  $71(1)$ &  $0.77(0.02)$ &  $92(1)$ &  $7.6(1.3)$ &  b \\
ESO027-001      & bdbar & $93(8)$ &  $0.22(0.06)$ &  $53(2)$ &  $0.65(0.05)$ &  $82(3)$ &  $12(3)$ &  \\
ESO027-008      & d     & $145(2)$ &  $0.64(0.02)$ &  N/A& N/A& N/A& N/A& b \\
\dots \\
UGC09951        & d     & $\sim120$ &  $\sim0.36$ &  N/A& N/A& N/A& N/A& c d \\
UGC10020        & d     & N/A   & N/A   & N/A   & N/A   & N/A   & N/A   & c f \\
UGC10437        & d     & $167(4)$ &  $0.18(0.03)$ &  N/A& N/A& N/A& N/A& c \\
UGC10445        & bd    & $124(3)$ &  $0.32(0.03)$ &  N/A& N/A& $98(2)$ &  N/A& \\
UGC12707        & dbar  & $43(3)$ &  $0.46(0.03)$ &  $65(2)$ &  $0.72(0.03)$ &  $86(2)$ &  $14(2)$ &  b \\
\hline
\end{tabular}
\caption{Galaxies in \sfg{} subsample. Column 1 is the galaxy name, column 2 indicates the model used in \textsc{DiskFit} (b=bulge, d=disk, and bar), column 3 is the position angle of the disk in degrees (numbers in parentheses are the uncertainties on the reported values), column 4 is the ellipticity of the disk $(1-b/a)$, column 5 is the position angle of the bar in degrees, column 6 is the bar ellipticity, column 7  is the fraction of the model light in the disk, column 8 is the fraction of the model light in the bar, and column 9 is the fit flag. The fit flags are a: Excluded galaxies with $|PA_{Disk} - PA_{Bar}| < 5^\circ$. b: Nucleus included in \textsc{galfit} fit, but not in \textsc{DiskFit} fit. c: Two disks fit in \textsc{galfit}, one disk fit in \textsc{DiskFit}. d: Galaxies for which the only model that we could fit was disk-only, with fewer surface brightness rings than used in other \textsc{DiskFit} fits. e: Galaxies for which we fit the same components as in the \textsc{galfit} fits, with fewer surface brightness rings than used in other \textsc{DiskFit} fits. f: Galaxies for which \textsc{galfit} parameters used as \textsc{DiskFit} inputs caused the \textsc{DiskFit} minimization to fail. The full table is shown in the digital appendix Table A1.\label{table1}}
\end{table*}
%\twocolumn

\section{Fitting of Synthetic Galaxies}\label{section3}
%Validating software on a relevant test sample can be very valuable when interpreting results, as it provides a control sample with known quantities. This can reveal whether any systematics in the real data are physical, or just an artifact of the fitting procedure. In this section, we describe our test sample (\S\ref{section3_1}) and present our results and their implications (\S \ref{section3_2}).
Validating the performance of a photometric decomposition algorithm on synthetic galaxies before proceeding to interpret fits to real data is essential for disentangling physical effects from software limitations. In this section we describe the simulated \sfg{}-like galaxy images that we use to test \textsc{DiskFit}'s performance (\S \ref{section3_1}), as well as the implications of our \textsc{DiskFit} model results on our fitting procedure for real galaxies (\S\ref{section3_2}).
\subsection{Setup and Fitting Procedures}\label{section3_1}
\noindent
We draw from the statistical properties of the subsample galaxies obtained from the fourth data reduction pipeline of \sfg{}   \citepalias{2015ApJS..219....4S} to create 400 images of synthetic galaxies, each containing a disk, a bar, and a bulge. The goal is to simulate galaxies with idealized structural components and image properties similar to the \sfg{} galaxy subsample in order to quantitatively assess \textsc{DiskFit}'s performance in this regime. For simplicity we adopt parametric surface brightness profiles for the synthetic galaxies, even though \textsc{DiskFit} recovers them non-parametrically.

We adopt exponential surface brightness profiles for the simulated disks:
\begin{equation}
\Sigma_{disk}(r) = \Sigma_{0,disk} e^{(-r/r_d)},
\end{equation}
with a disk scale length $r_d$, and a central surface brightness $\Sigma_{0,disk}$. The simulated surface brightness distribution of the bar is given by the Ferrers function \citep{1877ferrers}:
\begin{equation}
\Sigma_{bar}(r) =
\begin{cases} 
\Sigma_{0,bar} \left[1- \left(r/r_{bar} \right)^{2}\right]^2 & \quad r < r_{bar} \\
0 & \quad r \geq r_{bar}
\end{cases}
\end{equation}
with a truncation radius $r_{bar}$, and a central surface brightness $\Sigma_{0,bar}$. Finally, we adopt a S\'ersic function \citep{1963BAAA....6...41S} for the simulated bulges:
\begin{equation}
	\Sigma_{bulge}(r) = \Sigma_{eff} ~ exp \left(-b_n \left[ \left(\frac{r}{r_{eff}}\right)^{1/n} -1 \right] \right),
\end{equation}
with a S\'ersic index $n$, and surface brightness $\Sigma_{eff}$ at effective radius $r_{eff}$. We use the approximation $b_n = 1.99 n  - 0.327$ \citep{1989woga.conf..208C}.

\begin{figure}
\includegraphics[width=\columnwidth]{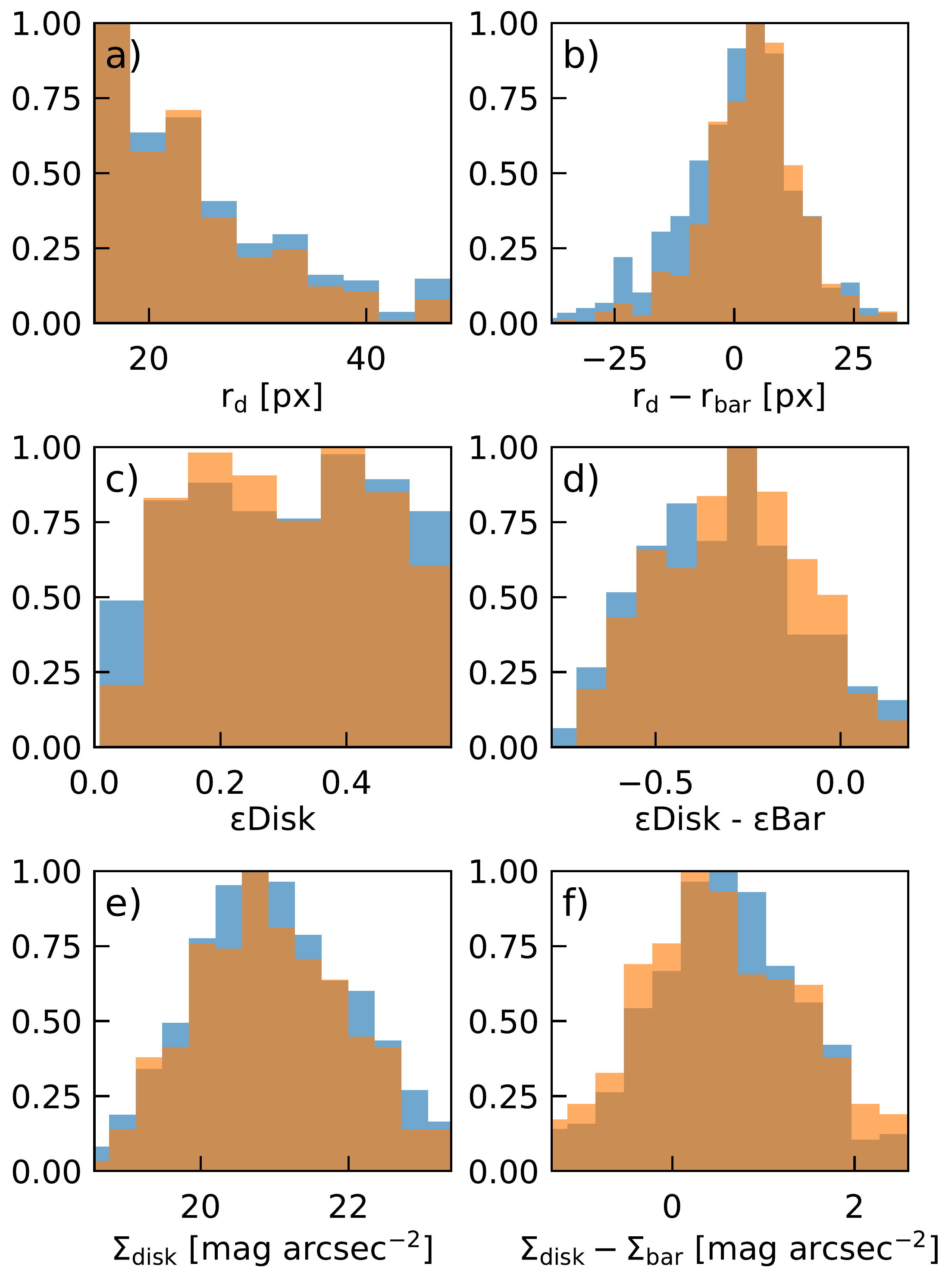}
	\caption{Distributions of: \textbf{a):} disk scale length, \textbf{b):} difference between disk scale length and bar radius, \textbf{c):} disk ellipticity, \textbf{d):} difference between disk and bar ellipticity, \textbf{e):} disk central surface brightness, \textbf{f):} difference between the central surface brightnesses of the disk and bar for the \sfg{}   subsample (blue) histograms and the simulated galaxy images (orange). \label{Figure2}}
\end{figure}

The component parameters in the simulated galaxy images are randomly drawn from their distributions in the \sfg{}   subsample as determined by \citetalias{2015ApJS..219....4S}, so that the real and synthetic samples span comparable ranges. The histograms in Figure \ref{Figure2} show the distribution of $r_d$, $\epsilon Disk$, $\Sigma_{disk}$, $r_{d}-r_{bar}$, $\epsilon Disk-\epsilon Bar$, and $\Sigma_{disk}-\Sigma_{bar}$ in the sample of synthetic galaxies. We also draw from the \sfg{} distributions for the bulge parameters $r_{eff}$, \eblg and $n$, imposing constraints on the relative surface brightnesses of the bulge and disk based on ranges from \citetalias{2015ApJS..219....4S}.

\begin{figure}
\centering
	\includegraphics[width=\columnwidth]{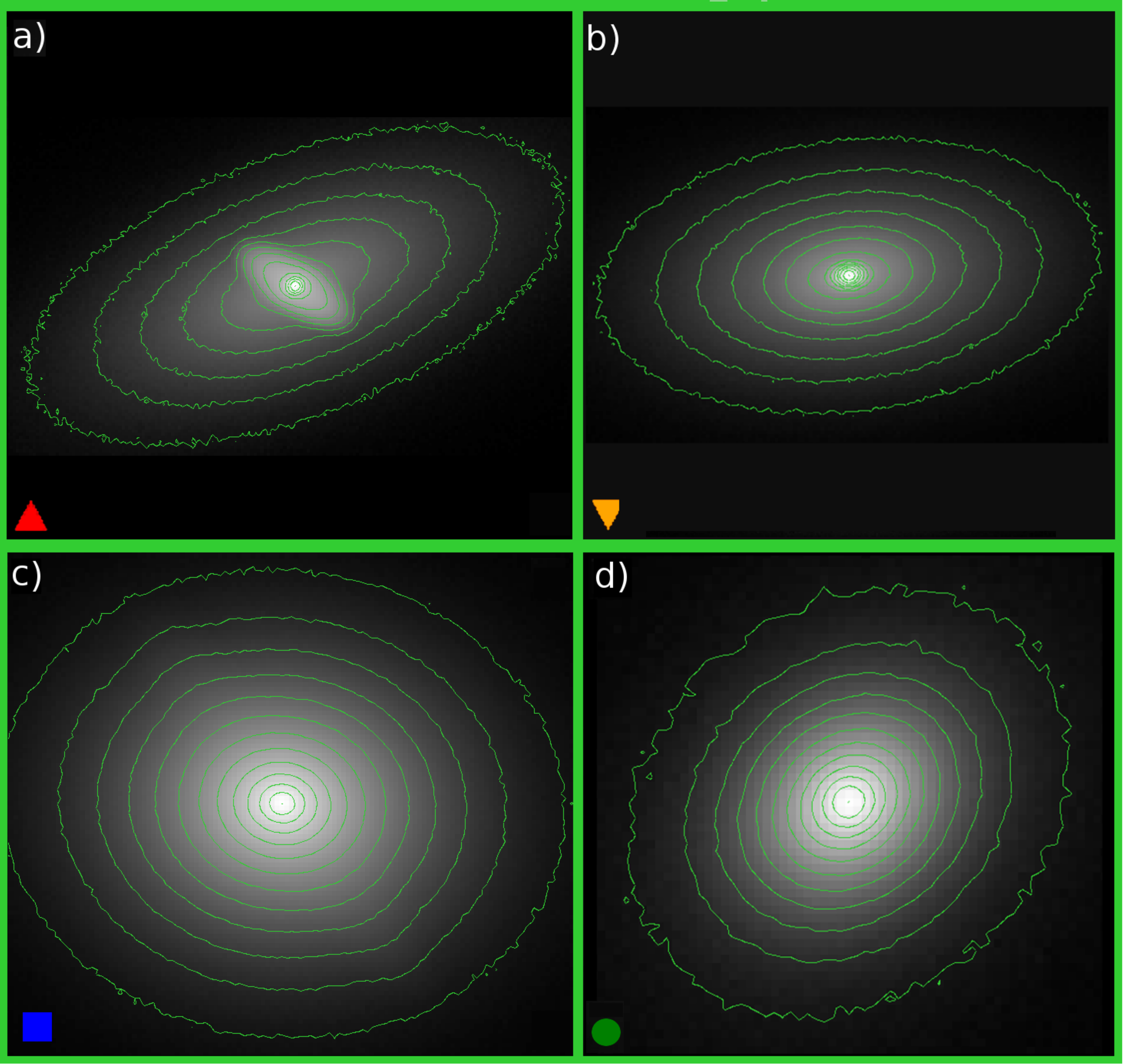}
	\caption{Simulated images of four synthetic galaxies. Each image is shown on a logarithmic scale, with 13 contour levels spaced by $\sim0.25$ dex. The location of these specific synthetic galaxies in Figure \ref{Figure5} is given by the symbol in the bottom-left corner of each panel in this figure, and the symbol colour indicates the best fitting surface brightness profiles in Figure \ref{Figure6}.\label{Figure3}}
\end{figure}

Once the structural properties of each galaxy are selected, the simulated images are made to be photometrically similar to those of the \sfg{}   subsample. We convolve the synthetic images using a Gaussian with full width at half maximum FWHM = 2.1 arcsec (with 0.75 arcsec/px), which closely resembles  the IRAC PSF \citep{2010PASP..122.1397S}. We add Poisson noise to the Gaussian-convolved images to mimic the source and sky background-dominated noise in the \sfg{}   images \citep{2010PASP..122.1397S}. We also create uncertainty maps for each galaxy using an algorithm similar to that adopted by \citetalias{2015ApJS..219....4S}. Examples of simulated galaxy images are shown in Figure \ref{Figure3}.

After the creation of the synthetic sample, we apply \textsc{DiskFit} to each simulated galaxy and its corresponding uncertainty map. We fit disk, bar, and bulge models for all systems. Input guesses for the disk PA, bar PA, $\epsilon Disk$, $\epsilon Bar$, S\'ersic $n$, \eblg and $r_{eff}$ are taken to be their true values in order to maximize the chances of recovery. We initially allow all parameters to vary freely in the models to determine the best fitting non-parametric surface brightness profiles for the disk and the bar. However, for reasons that we discuss in \S \ref{section3_2} we leave S\'ersic $n$ fixed to the true value for all figures following Figure \ref{Figure4}. We set \textsc{DiskFit} to correct for seeing using the same Gaussian PSF as adopted in the simulations. We estimate uncertainties on the fitted parameters using 20 bootstrap-resampled realizations of each model using the radial-rescaling method of \citet{2010MNRAS.404.1733S}, where the number of bootstrap realizations is limited by the computing resources available through the Centre for Advanced Computing (CAC) at Queen's University. The next section presents the results of these fits. 
\begin{figure*}
\includegraphics[width=\columnwidth]{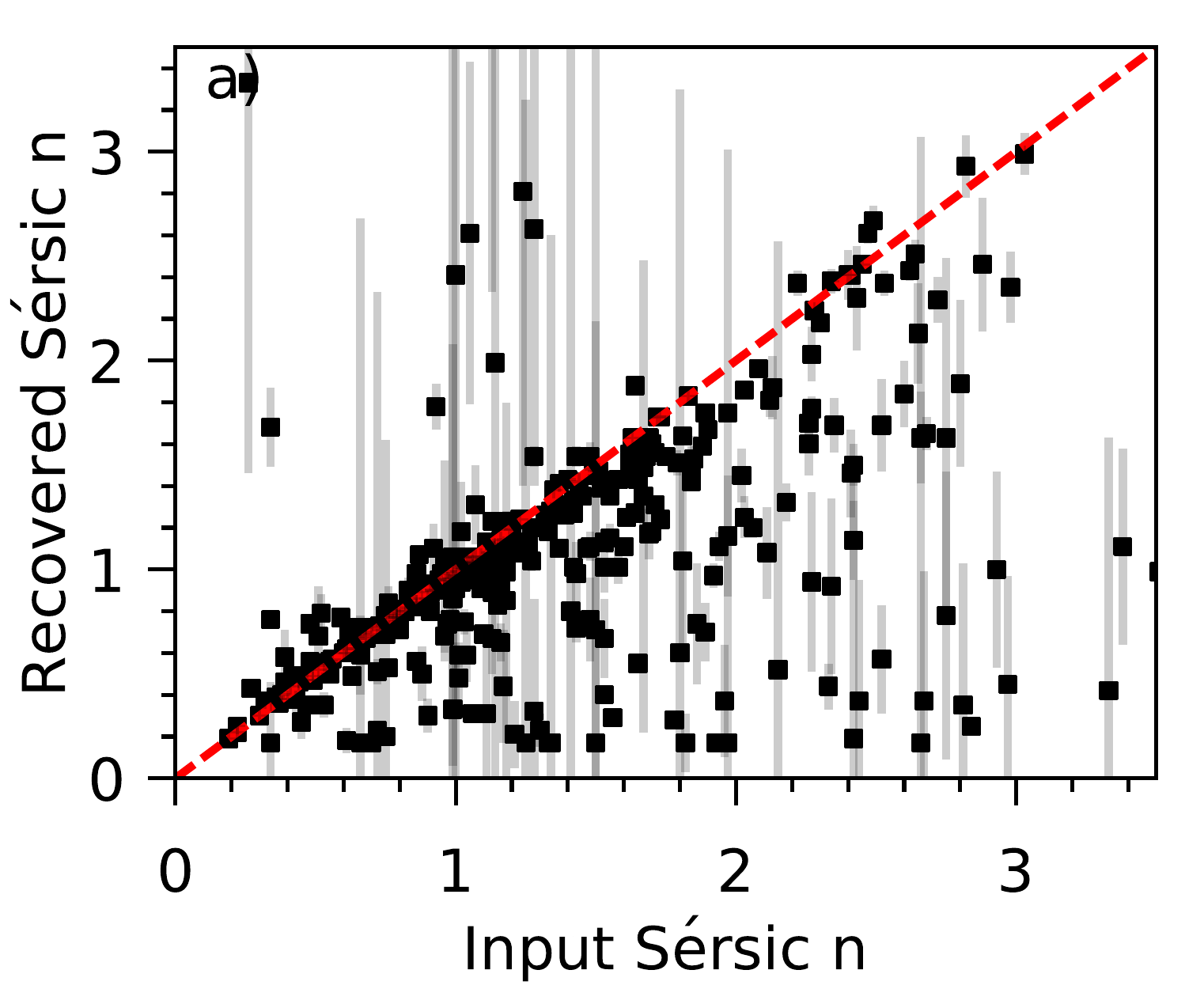}
\includegraphics[width=\columnwidth]{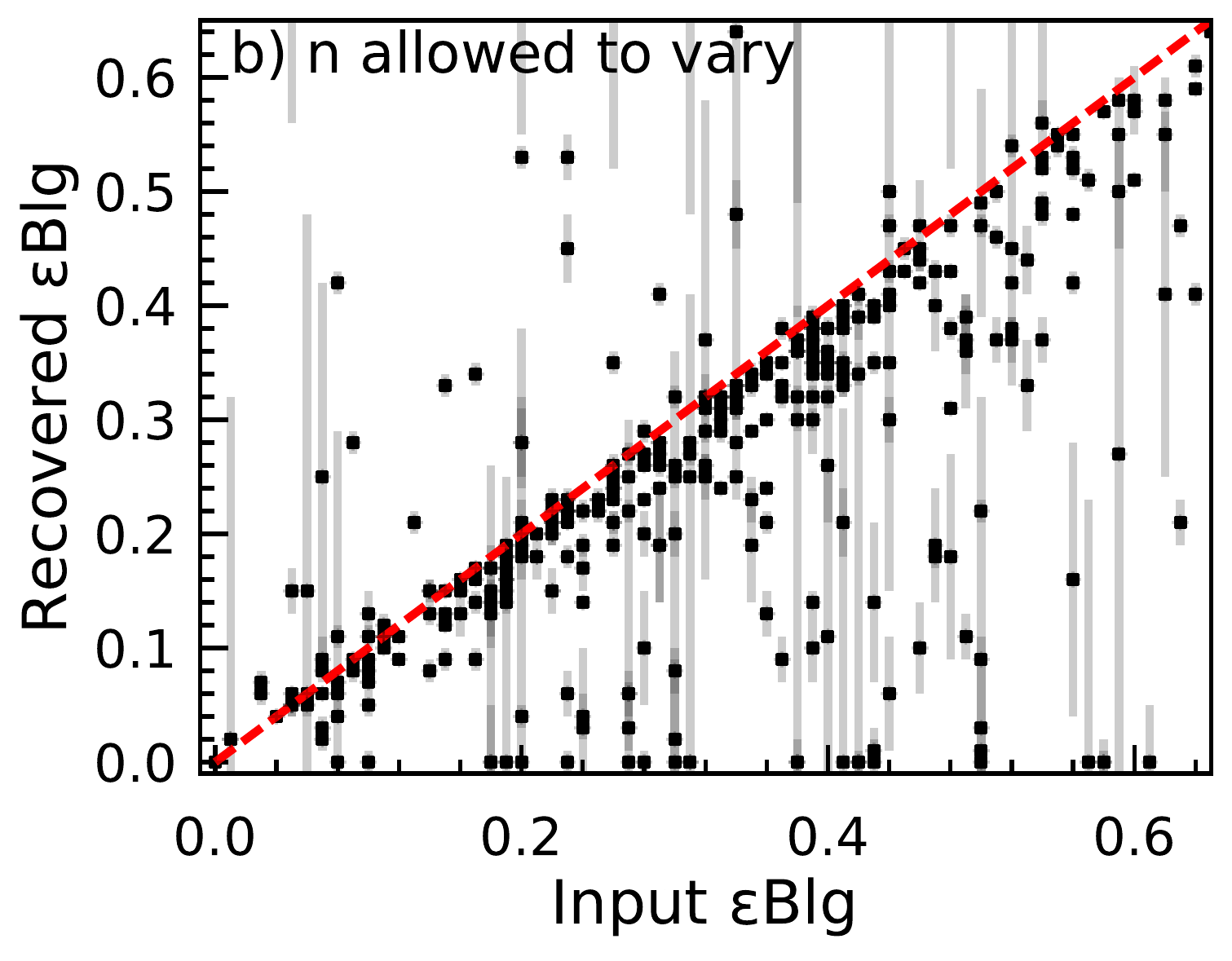}\includegraphics[width=\columnwidth]{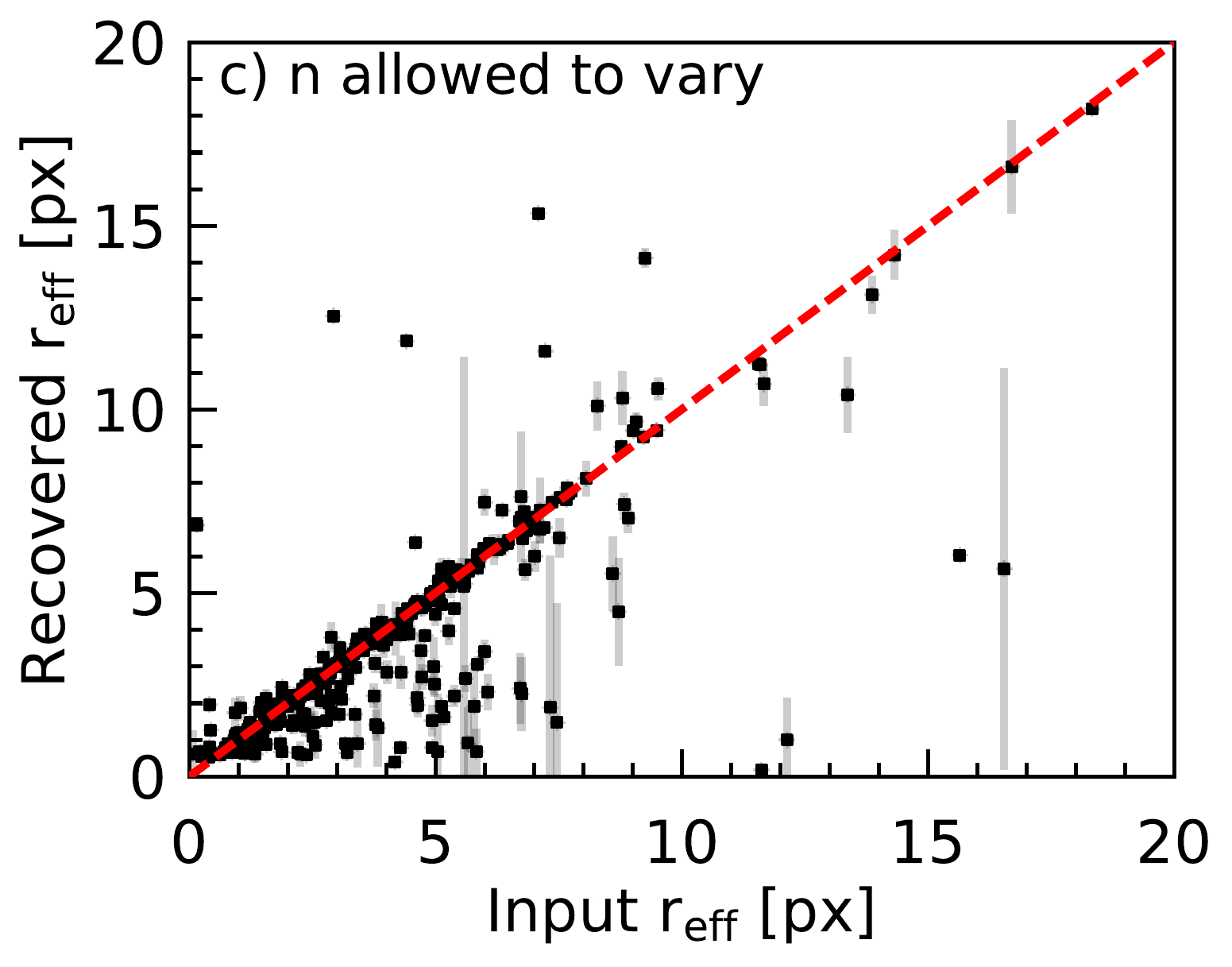}
\includegraphics[width=\columnwidth]{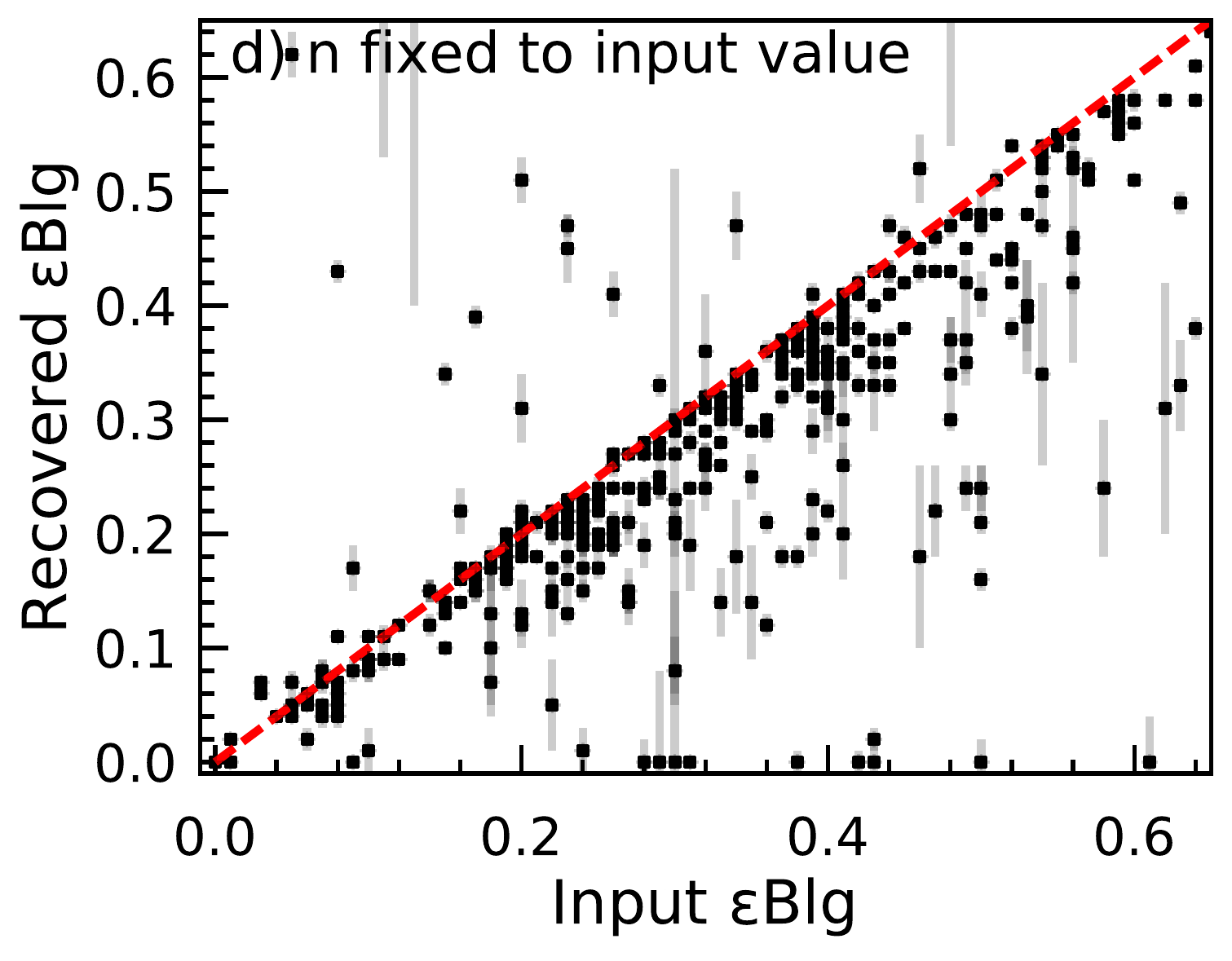}\includegraphics[width=\columnwidth]{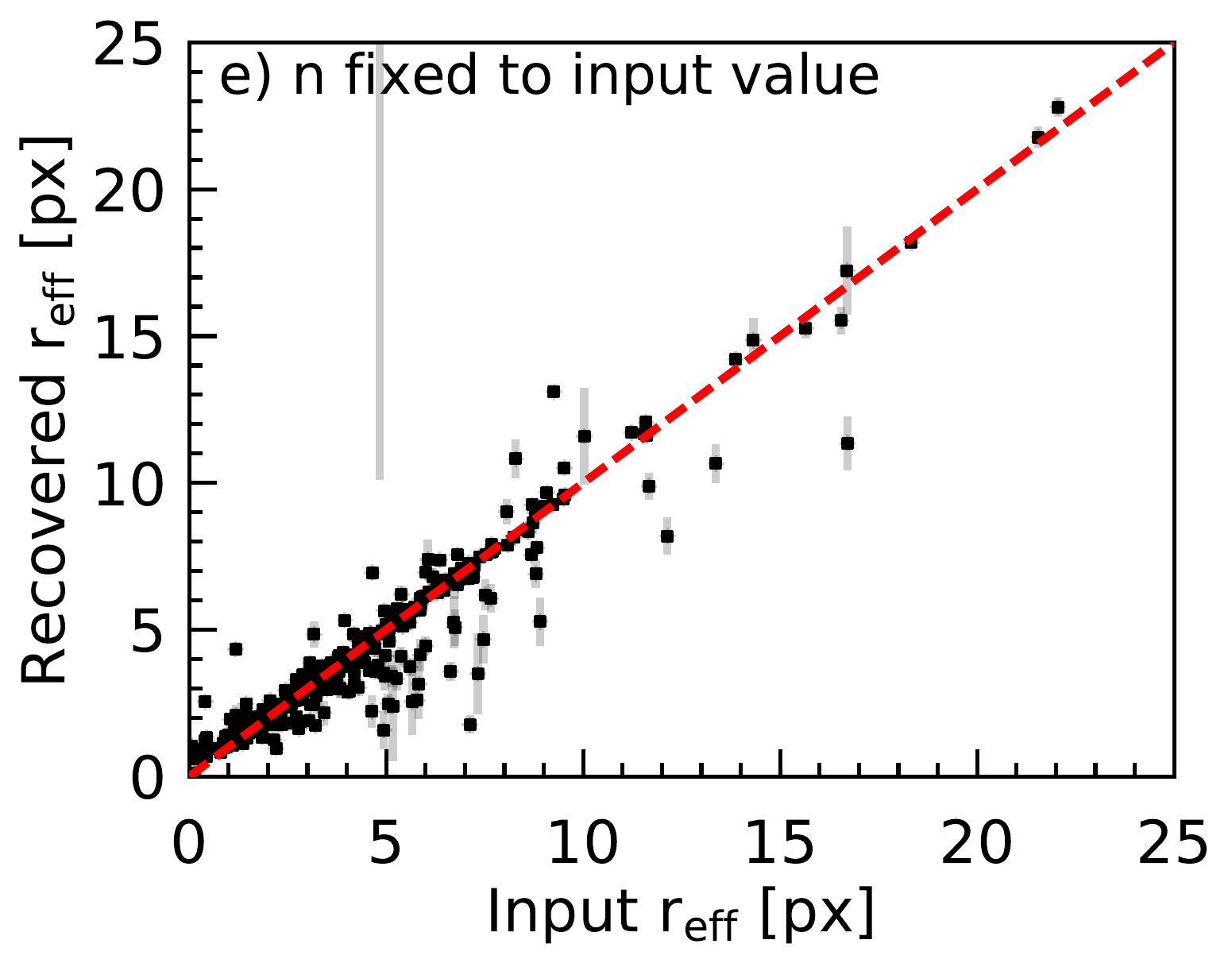}
	\caption{Recovered vs input bulge properties from \textsc{DiskFit} fits to synthetic galaxies: \textbf{a)}: S\'ersic $n$, \textbf{b),d)}: bulge ellipticity, \textbf{c),e)}: bulge effective radius. In \textbf{b)} and \textbf{c)}, S\'ersic $n$ is allowed to vary, while in \textbf{d)} and \textbf{e)} $n$ is fixed to the true value. In each plot, the dashed red line is the 1:1 relationship. \label{Figure4}}
\end{figure*}

\begin{figure*}
\includegraphics[width=\columnwidth]{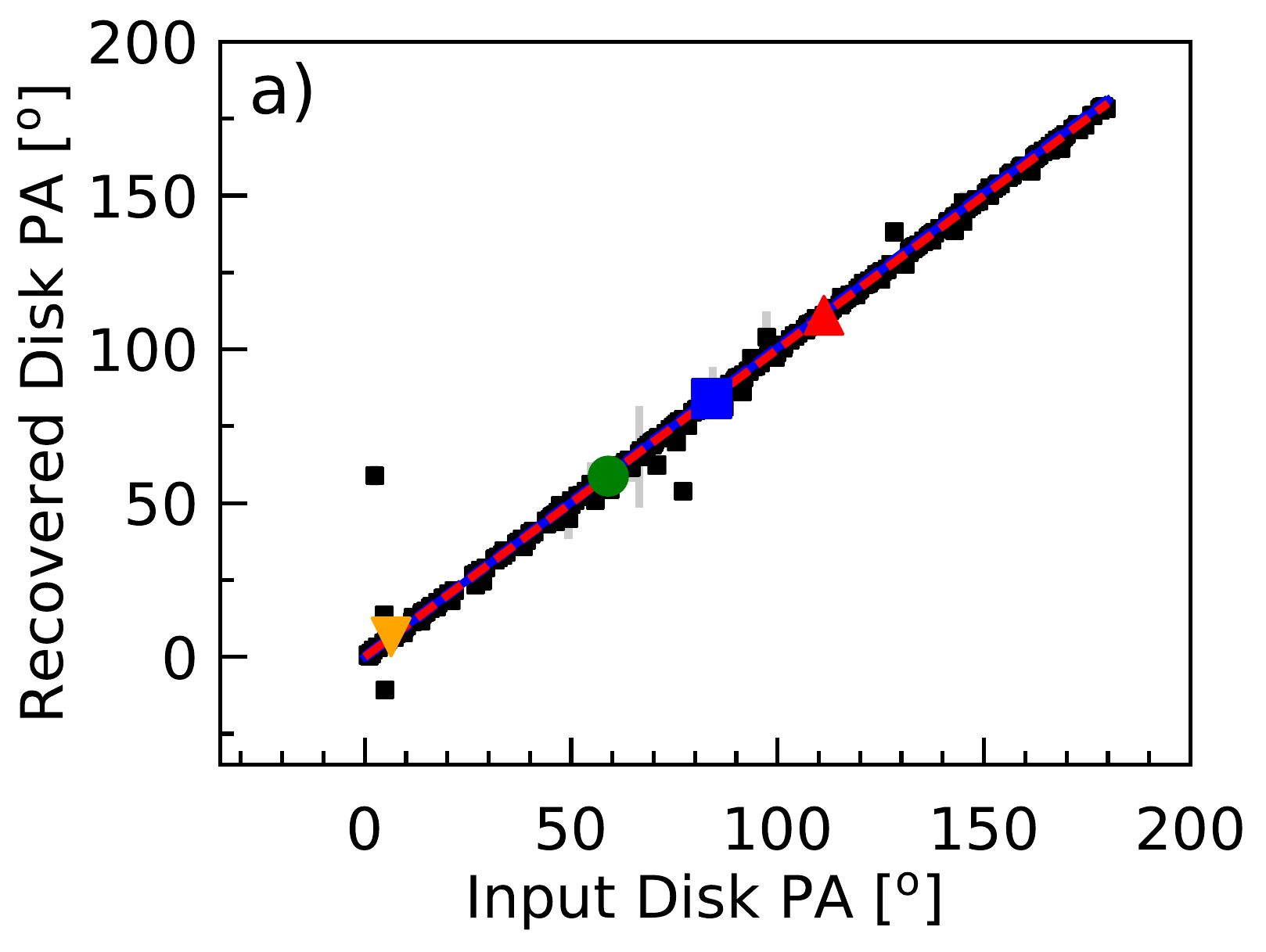}\includegraphics[width=\columnwidth]{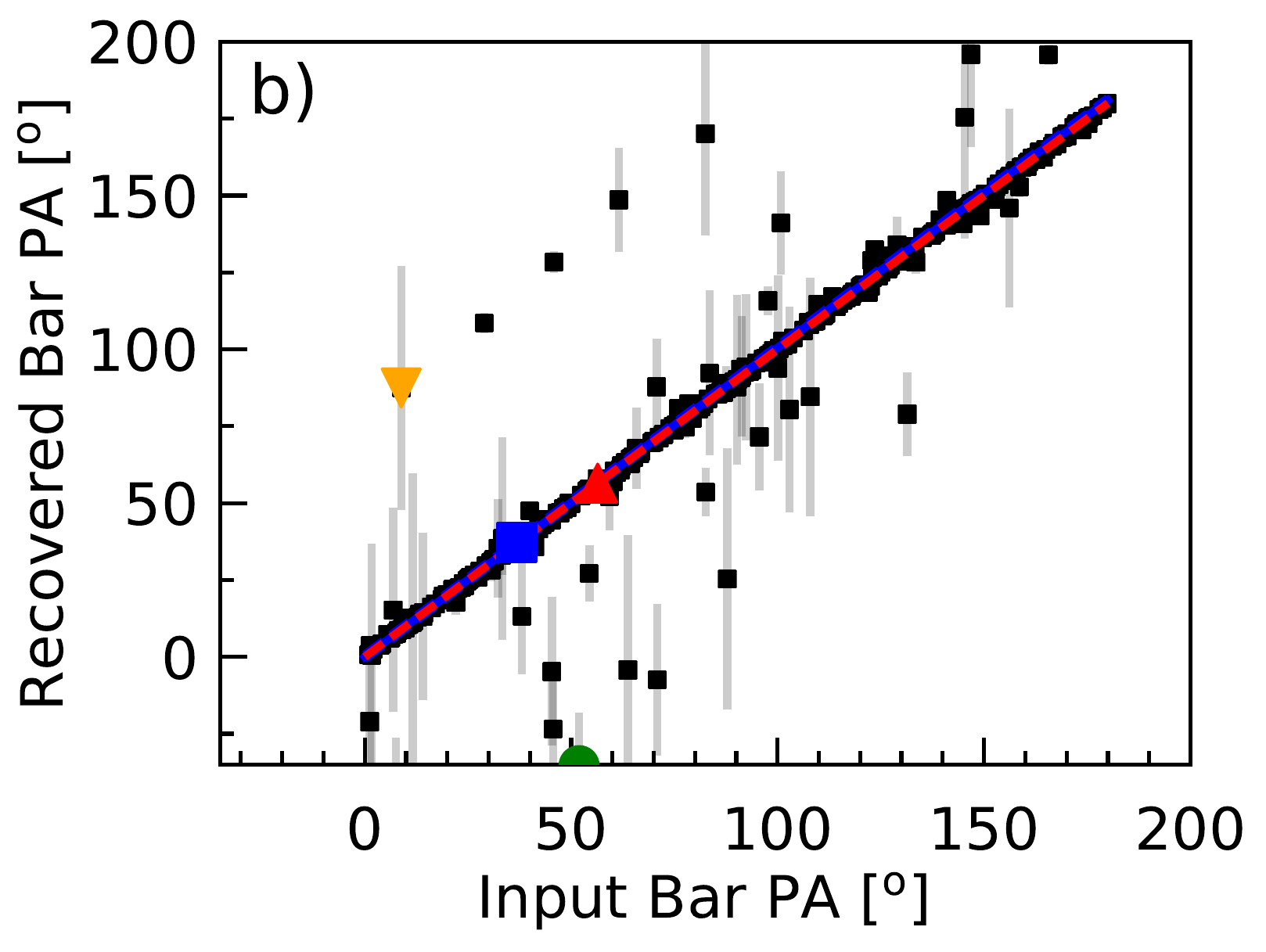}
\includegraphics[width=\columnwidth]{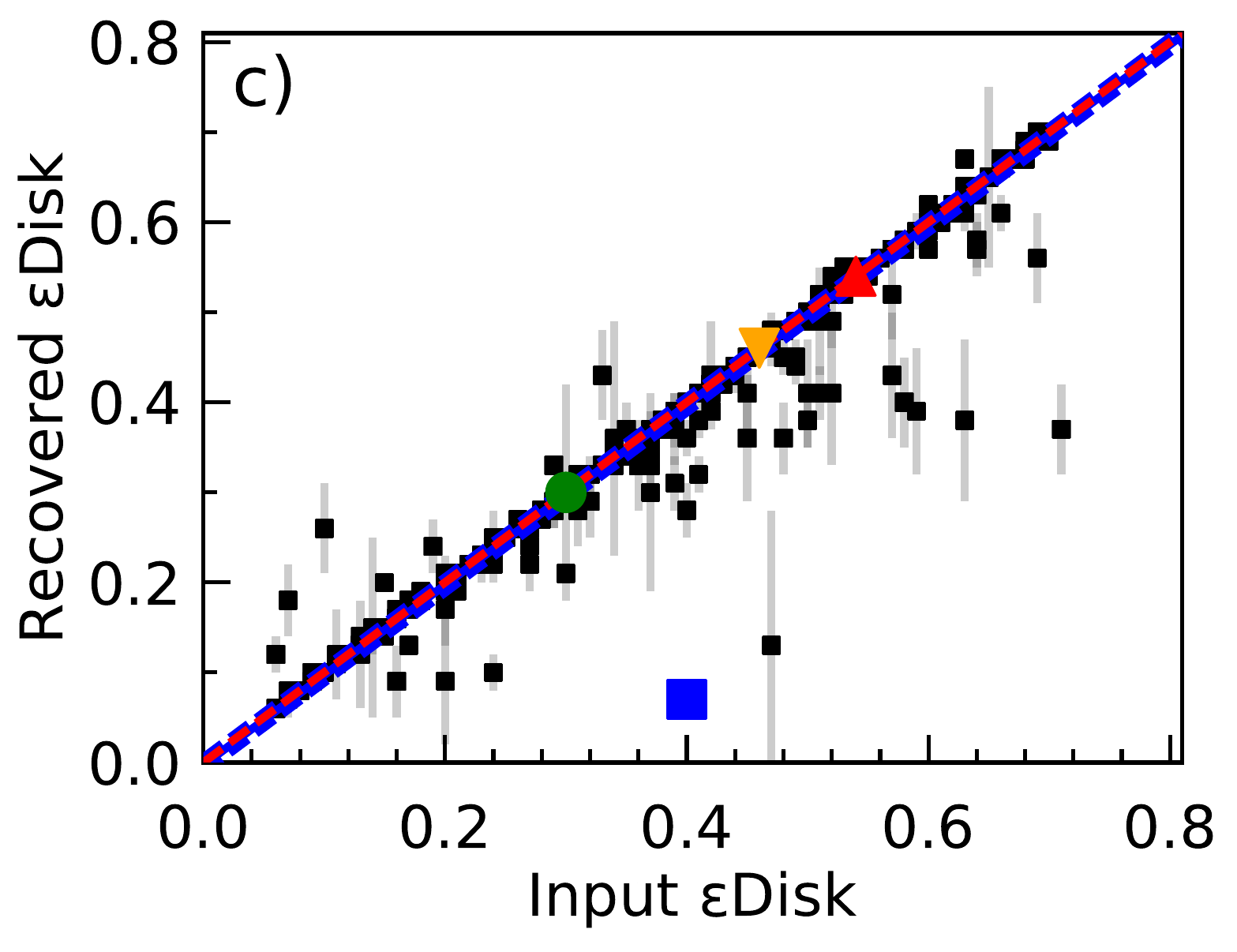}\includegraphics[width=\columnwidth]{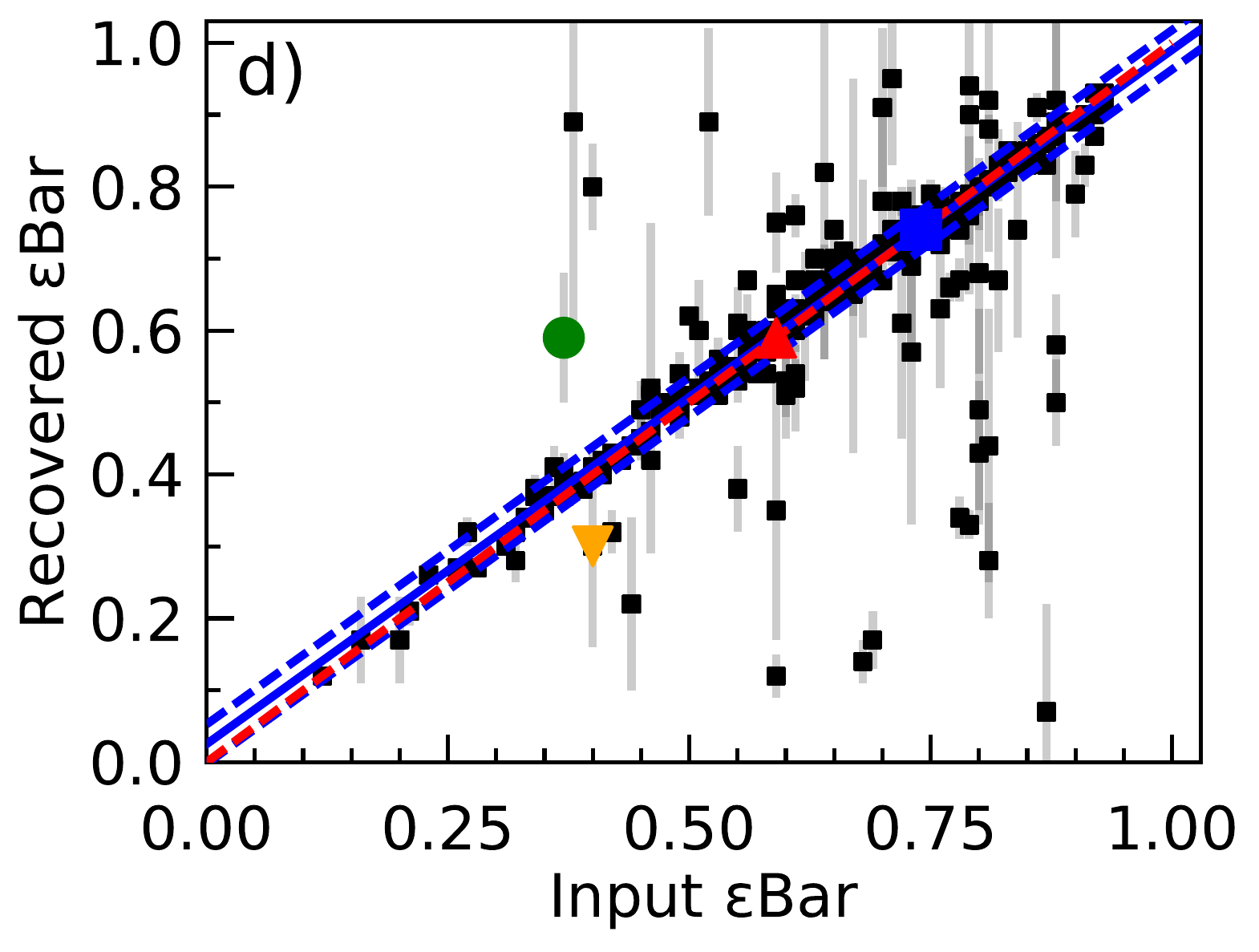}
	\caption{Recovered vs input disk and bar from \textsc{DiskFit} fits to synthetic galaxies: \textbf{a):} disk position angle, \textbf{b):} bar position angle, \textbf{c):} disk ellipticity, \textbf{d):} bar ellipticity. In each panel, the solid blue line is the best-fitting linear least squares relationship, the dashed blue lines show two times the normalized median absolute deviation (MAD/1.4826) about the best fit, and the dashed red line is the 1:1 relationship. The coloured points indicate the synthetic galaxy images shown in Figure \ref{Figure3}. \label{Figure5}}
\end{figure*}

\subsection{Results of Synthetic Galaxy Fits}\label{section3_2}
Figure \ref{Figure4} shows the relationship between the input and recovered bulge properties for the synthetic galaxy images. Panels a), b) and c) show the bulge properties recovered when all bulge parameters are allowed to vary in the fits. It is clear that S\'ersic $n$ is not reliably recovered, and that there is significant scatter between the input and recovered \eblg and $r_{eff}$. The situation improves when S\'ersic $n$ is fixed to the input value, as shown in panels d) and e) in Figure \ref{Figure4}. The recovered values for $r_{eff}$ are scattered symmetrically about the 1:1 relation. The recovered \eblg loosely follows the 1:1 line, with the majority of recovered values underestimating the true ones: this is a consequence of \textsc{DiskFit} assigning bulge light to the (non-parametric) disk. The fraction of light in the bulge (a cumulative measure computed from n, \eblg and $r_{eff}$) is not recovered by DiskFit regardless of whether or not n is held fixed.

Our simulations imply that \textsc{DiskFit} cannot reliably recover S\'ersic $n$ in S$^4$G-like imaging, i.e simulated galaxy images in which the range of structural parameter values, the resolution, and the image noise resemble that of \sfg{}. We therefore fix this parameter in all subsequent \textsc{DiskFit} models of both simulated and real systems, and have used our simulations to verify that the disk and bar properties are unaffected by this choice beyond the bulge region. Figures \ref{Figure4}d) and \ref{Figure4}e) imply that, even with $n$ fixed to the correct value, the other S\'ersic bulge properties are not well-recovered. It is therefore likely that the \textsc{DiskFit} models of the \sfg{}   subsample (where the true $n$ is unknown) are unreliable in the bulge region of each galaxy, and we do not consider this region further here.   All subsequent models of synthetic and real galaxies keep $n$ fixed.

A comparison between the input and recovered position angles and ellipticities of the disk and bar in our synthetic galaxy images is shown in Figure \ref{Figure5} and the best fitting values from least-squares linear fits to each trend are given in Table \ref{table_stats}. Panels a) and b) of Figure \ref{Figure5} show that for the vast majority of simulated galaxies, the disk and bar PAs are accurately and precisely recovered by \textsc{DiskFit}. We note that synthetic galaxies with outlying values of the recovered disk PA are mostly low-inclination systems with $\epsilon Disk < 0.1$. The relationships between the input and recovered disk and bar ellipticities in Figures \ref{Figure5}c) and \ref{Figure5}d) show comparatively more scatter, but the correlations remain strong. Table \ref{table_stats} shows that the median absolute deviation normalized to equal the standard deviation for Gaussian distributions (MAD/1.4826) about the best-fitting linear relationship implies an extremely small scatter of the best fitting values from their true ones for most synthetic galaxy fits.

Insight into the nature of the outliers can be gleaned by examining the coloured points in Figure \ref{Figure5}, which correspond to the individual galaxies in Figure \ref{Figure3}. The red triangles in Figure \ref{Figure5} show an example of a synthetic galaxy, highlighted in Figure \ref{Figure3}a), where the recovered properties match the input values within the uncertainties; this system has a relatively bright bar ($r_{bar}= $ 23 px, where the bar comprises 62\% of the total galaxy light) at a distinct position angle from the underlying disk.

By contrast, the outlying green circles in Figure \ref{Figure5}b) and \ref{Figure5}d) illustrate \textsc{DiskFit}'s higher likelihood of failing to recover a weak bar ($r_{bar}\simeq $ 16 px with 0.3\% galaxy light) in the synthetic galaxy in Figure \ref{Figure3}d); this is perhaps to be expected, since the bar isn't visible by eye.  \textsc{DiskFit} fares better at recovering the longer and slightly stronger bar ($r_{bar}= $ 56 px with  0.9\% galaxy light) in the synthetic image in Figure \ref{Figure3}c), represented by the blue squares in Figure \ref{Figure5}; the bar properties are well-recovered, though $\epsilon Disk$ is strongly under-estimated. These individual examples illustrate that synthetic galaxies with recovered properties that deviate from the input ones tend to have one structural component that is much brighter than the others.

The yellow triangles in Figure \ref{Figure5}, which show the recovered parameters of the synthetic galaxy in Figure \ref{Figure3}b), illustrate how \textsc{DiskFit} performs when the disk and bar PA are closely aligned (the two have identical position angles in this specific case). The Disk PA and $\epsilon$Disk are well-recovered, but $\epsilon_{bar}$ is poorly recovered and the Bar PA is completely lost. 

\begin{figure}
	\includegraphics[width=\columnwidth]{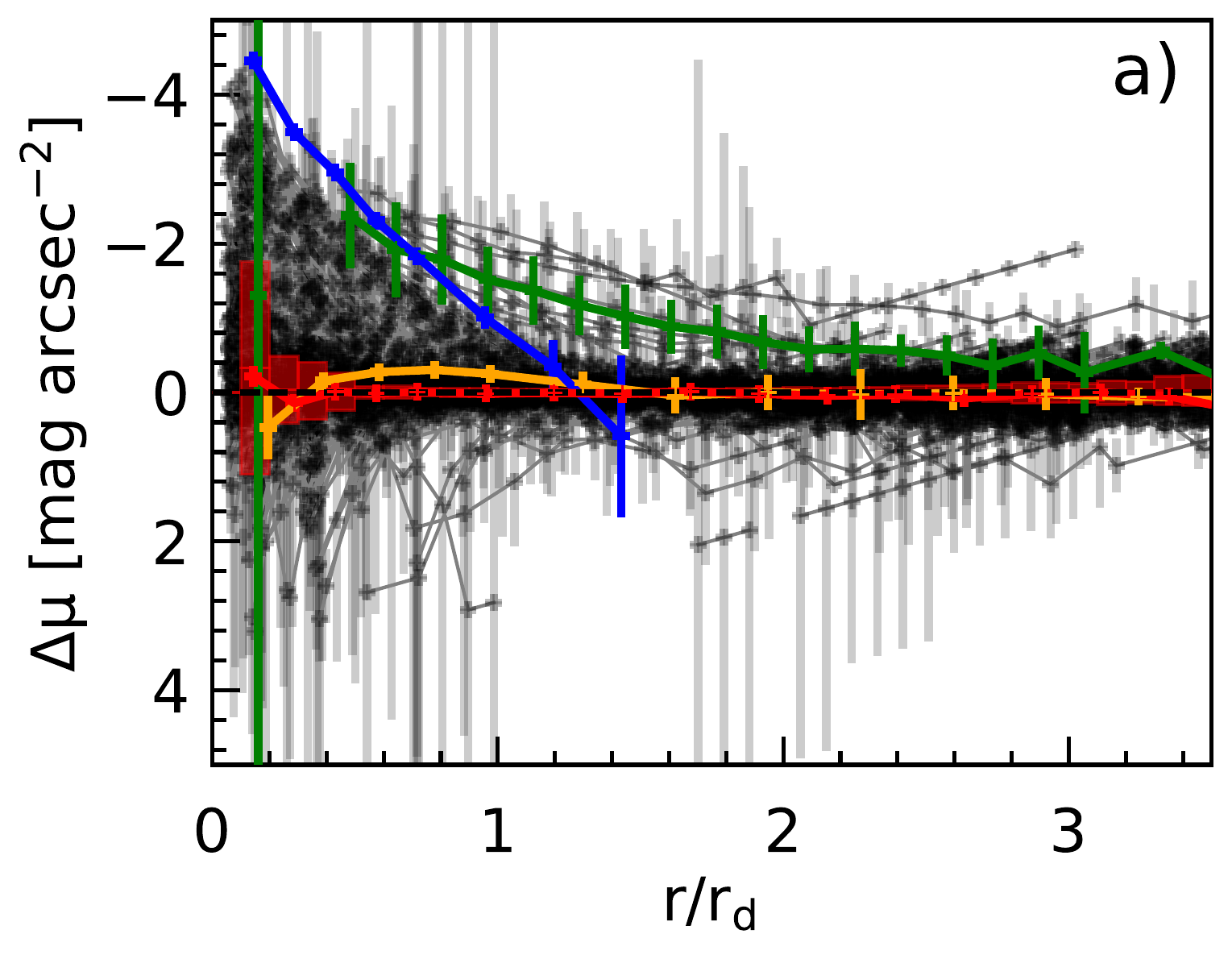}
	\includegraphics[width=\columnwidth]{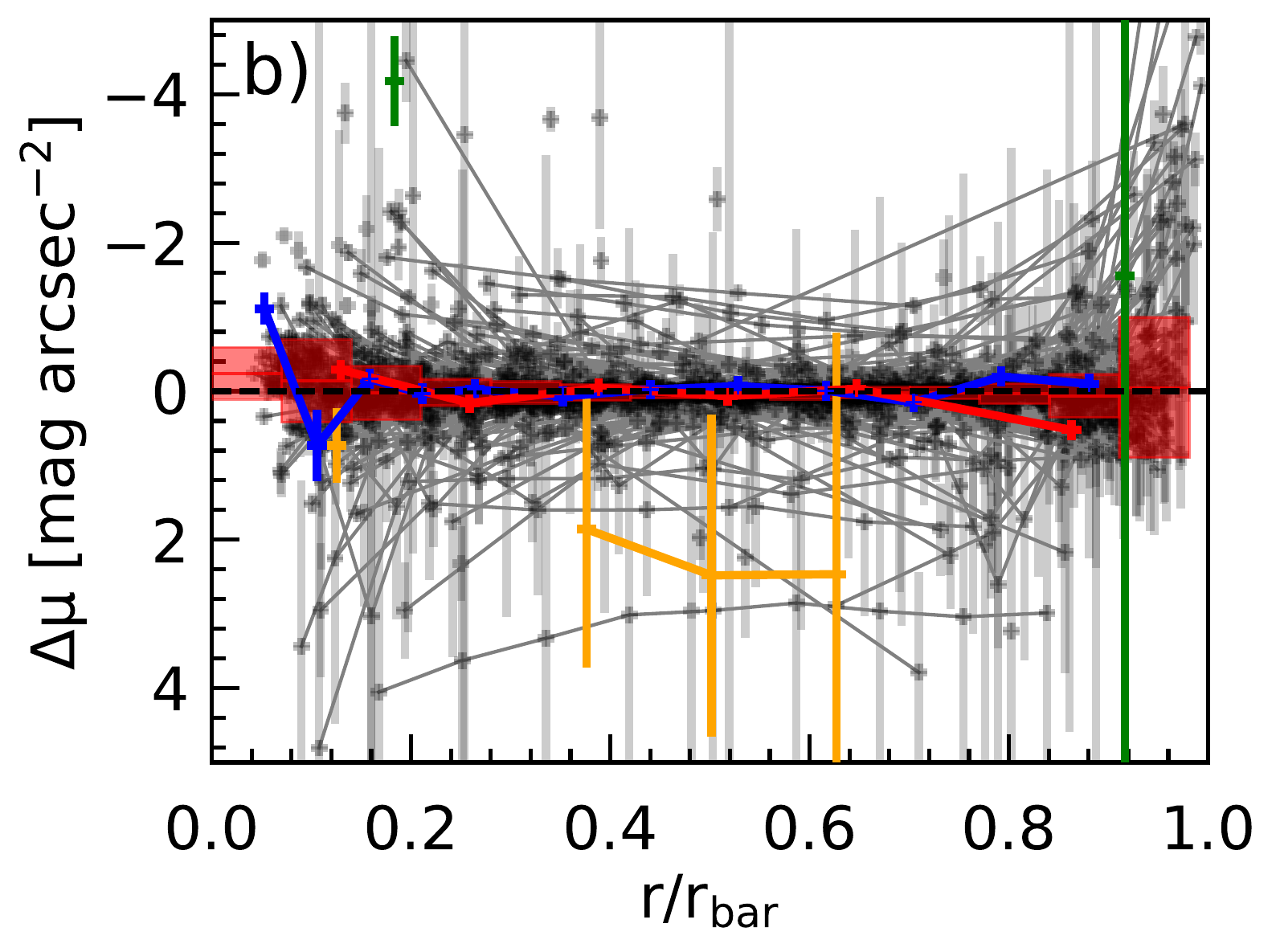}
	\caption{Difference between the recovered (non-parametric) and input (parametric) surface brightness profiles of \textbf{a):} the disk and \textbf{b):} the bar measured by \textsc{DiskFit} from the simulated galaxy images, where $\Delta_{\mu} = \mu_{recovered} - \mu_{input}$. Connected points belong to the same galaxy, and the radial range has been normalized to the input $r_{d}$ and $r_{bar}$, respectively. The dashed black line indicates $\Delta_{\mu}=0$, and red rectangles indicate the interquartile range. The coloured lines correspond to the best-fitting profiles to the synthetic galaxy images with same-coloured symbols in  Figure \ref{Figure3}\label{Figure6}.}
\end{figure}

Figure \ref{Figure6} compares the surface brightness profiles of the disk and the bar recovered by \textsc{DiskFit} for the synthetic galaxies, where $\Delta \mu = \mu_{recovered} - \mu_{input}$ at the discrete radii selected during modelling. Points with $\Delta \mu > 0$ therefore correspond to an overestimation of the light profile by \textsc{DiskFit}, and points with $\Delta \mu < 0$ correspond to an underestimation. We normalize the radial axes by $r_d$ and $r_{bar}$ of the (parametrically-constructed) input synthetic galaxies for convenience, but emphasize that these quantities have no meaning in the \textsc{DiskFit} context since the disk and bar are modelled non-parametrically.

The narrow interquartile range of $\Delta \mu$ (red shaded regions) in Figure \ref{Figure6}a) shows that in general, for $0.5 < r/r_{d} < 3$, exponential disk surface brightness profiles are reliably recovered non-parametrically by \textsc{DiskFit}.  At smaller $r$, the disk and bar surface brightnesses are not well-recovered due to confusion with the bulge (which itself isn't reliably modelled; see Figure \ref{Figure4}). Note that the extreme outliers here consist of galaxies similar to those shown in Figure \ref{Figure3}b) - \ref{Figure3}d) (coloured yellow, blue, and green in Figures \ref{Figure5} and \ref{Figure6} respectively), where the disk-bar degeneracies are hardest to break. Figure \ref{Figure6}b) shows that the bar is also reliably recovered by \textsc{DiskFit} in the range $0.2 < r/r_{bar} < 0.8$. Note that because \textsc{DiskFit} models the surface brightness profiles of the disk and bar non-parametrically, the two components become degenerate in the model when their position angles are aligned and their ellipticities are similar. This is clearly illustrated by the yellow lines in Figure \ref{Figure6}, which show the best-fitting profiles corresponding to the synthetic galaxy in Figure \ref{Figure3}b). Our simulations suggest that, for intermediate-inclination disks, \textsc{DiskFit} is unreliable at distinguishing the disk and bar components when $\Delta PA = | PA_{disk} - PA_{bar} | \lesssim 5^\circ$ (see \citealt{2015MNRAS.451.4397H} and \citealt{2016A&A...594A..86R} for similar findings in kinematic \textsc{DiskFit} models). Finally, for $r/r_{bar} > 0.85$, the noise in \textsc{DiskFit}'s recovery of the bar increases as the bar light fades sharply (in the Ferrers profile, the bar fades by $\sim 7$ mag arcsec$^{-2}$ for $0.85 < r/r_{bar} < 1$). 

The outcome of our \textsc{DiskFit} modelling of the simulated galaxies has important implications for our models of the \sfg{}   subsample described in \S \ref{section4_1}. We find that bulge properties are not reliably modelled. However, the position angles, ellipticities and surface brightness profiles of galaxies with exponential disks and relatively long, relatively bright Ferrers bars (usually those at least barely distinguishable by eye) are well-recovered beyond the bulge region. The notable exception to these criteria are systems where the disk and bar position angles lie within 5$^\circ$ of one another, which may cause degeneracies in \textsc{DiskFit}'s non-parametric models.

\section{Fitting of \sfg{}   Galaxies}\label{section4}
We now proceed to model the \sfg{}   subsample listed in Table \ref{table1} with \textsc{DiskFit}, in order to compare the resulting components with those obtained from the human-supervised application of \textsc{galfit} by \citetalias{2015ApJS..219....4S}. We use the synthetic galaxy models of \S\ref{section3} to guide the selection of subsample galaxies that we attempt to fit (\S\ref{section4_1}), as well as to interpret our results (\S\ref{section4_2}).

\subsection{Setup and Fitting Procedures}\label{section4_1}
 For each system in our subsample, we fit for the same combination of disk, bar, and/or bulge as included in the final \textsc{galfit} decomposition presented by \citetalias{2015ApJS..219....4S}. The components included in each of the models are listed in column 2 of Table \ref{table1}. 

We attempt to model most galaxies in the subsample with \textsc{DiskFit}, using the disk geometry ($\epsilon Disk$, Disk PA) obtained by \citetalias{2015ApJS..219....4S} as input guesses, as well as the \citetalias{2015ApJS..219....4S} bar geometry ($\epsilon Bar$, Bar PA) when one is included.  We use the exponential disk scale length $r_d$ estimated by \citetalias{2015ApJS..219....4S} to define the radii at which we solve for the surface brightness of each component, which we set to every third pixel from $r=0$ to $r=3 r_d$, and then every fifth pixel out to $6r_d$. The best-fitting model values do not depend strongly on the adopted radial sampling cadence. We estimate uncertainties on the model parameters using 200 bootstrap realizations, a number again constrained by the computing resources available to us through the CAC. As discussed in \S \ref{section3_2} we will not compare models of the bulge because our simulations imply that \textsc{DiskFit} is unreliable in that region. For models including a bulge, we fix S\'ersic $n$ to the \citetalias{2015ApJS..219....4S} value and allow for $r_{eff}$ and \eblg to vary, using the best-fitting values from \citetalias{2015ApJS..219....4S} as \textsc{DiskFit} initial guesses. Parameters listed as `N/A' in Table \ref{table1} were not included in the model for that particular galaxy. 

The results of the fits for each galaxy in the \sfg{}   subsample are given in Table \ref{table1}, and the data files containing the resulting surface brightness profiles are available in the electronic version of the paper. 

The simulations of \S \ref{section3_2} imply galaxies with |Disk PA - Bar PA| $\lesssim5^\circ$ are not reliably modelled by \textsc{DiskFit}. We preemptively exclude the 38 \sfg{} subsample galaxies for which this is the case in the \citetalias{2015ApJS..219....4S} models, listing them in Table \ref{table1} with flag `a'. We therefore attempt to fit a total of 532 galaxies. We note that some \textsc{galfit} models include a nuclear component (144 galaxies) or a second disk (89 galaxies), which we do not include in the \textsc{DiskFit} models; these systems are flagged with a `b' or `c' in the last column of Table \ref{table1}, respectively.

%suggests that particular morphologies are not optimally fit by \textsc{DiskFit}. We look for cases of real galaxies that have similar position angles of the disk and bar (|Disk PA - Bar PA| $<5^o$), and we preemptively exclude these in order to exclude fits that will likely be inaccurate. These 38 galaxies are listed 

Additionally, there are a few galaxies for which we struggled to recover the same components as those found by \citetalias{2015ApJS..219....4S} using \textsc{galfit}, with \textsc{DiskFit} getting `hung up' in parameter space during the minimization or after the first few bootstraps. We are able to recover the \textsc{galfit} morphology for some of these systems when we reduce the number of rings used in the \textsc{DiskFit} surface brightness profile to every fifth pixel from $r=0$ to $r=3 r_d$, and then every tenth pixel out to $6r_d$. We give these 19 galaxies an `e' flag in Table \ref{table1}. For other galaxies that we struggle to fit with the same process, we are able to fit a disk-only model (reporting approximate values), again using fewer surface brightness rings. These 8 galaxies are listed in Table \ref{table1} with flag `d'. Finally, there are 16 galaxies for which we were not able to use \textsc{galfit} parameters as input at all, with \textsc{DiskFit} failing to minimize for even disk-only models. We flag these galaxies with an `f' in Table \ref{table1}, and return to them in \S\ref{section5}. We omit the \sfg{} subsample galaxies with flags `a', `d', `e', and `f' from our comparisons with the results of \citetalias{2015ApJS..219....4S}. This `comparison sample' includes 489 members. %Representative fits to individual systems in this sample are shown in Figures \ref{Figure7}-\ref{Figure9}, and fit results for the sample as a whole are shown in Figures \ref{Figure10}-\ref{Figure11}.

\begin{figure}
	\centering
	\includegraphics[width=0.9\columnwidth]{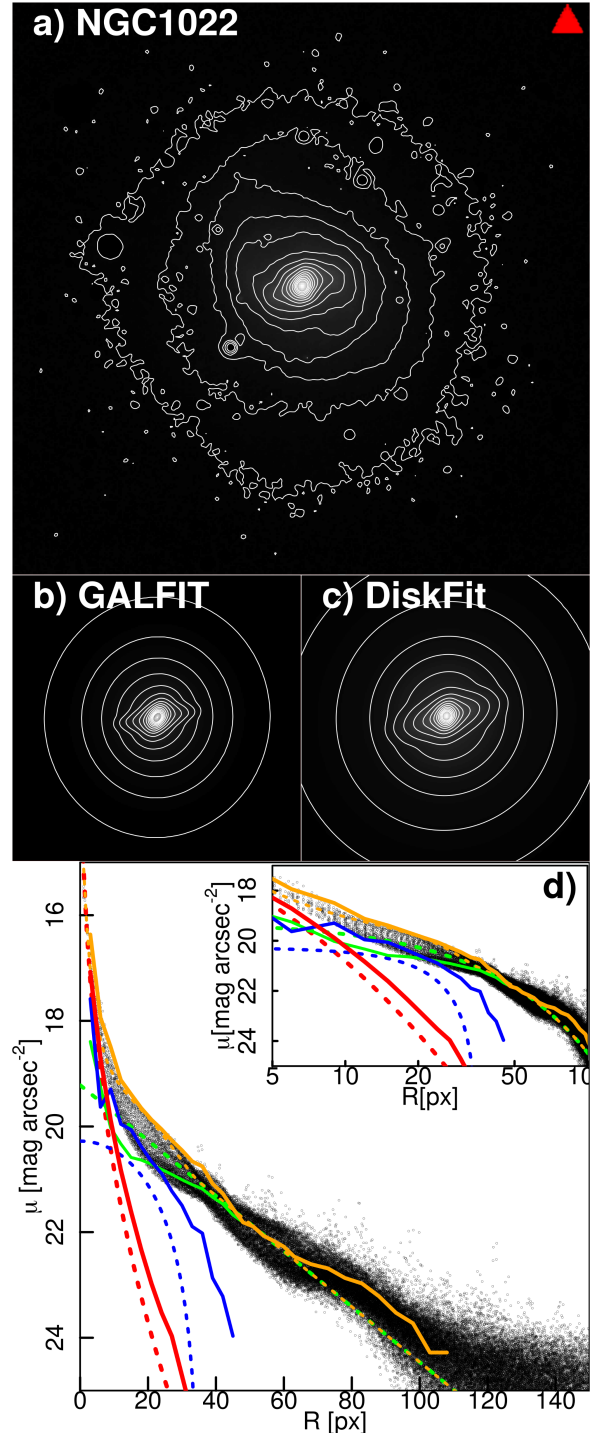}
	\caption{Images of \textbf{a):} NGC1022 from \sfg{}, \textbf{b):} the \citetalias{2015ApJS..219....4S} \textsc{galfit} model of NGC1022, and \textbf{c):} the \textsc{DiskFit} model of NGC1022 on the same angular scale. There are 20 contours separated by $\sim$ 0.16 dex. In the surface brightness profiles in \textbf{d)}, the bulge model is shown in red, the bar model is shown in blue, the disk model is shown in green, and the full model is shown in yellow. The \textsc{DiskFit} fit is shown with solid lines and the \textsc{galfit} fit is shown with dashed lines. Both the \textsc{DiskFit} and \textsc{galfit}  curves show surface brightness as a function of isophotal semi-major axis. The black dots show the surface brightness vs. sky distance from the centre of the image. NGC1022 is an example of a comparison sample galaxy with similar \textsc{DiskFit} and \textsc{galfit} fits.}
   \label{Figure7}
\end{figure}
\begin{figure}
	\centering
	\includegraphics[width=0.9\columnwidth]{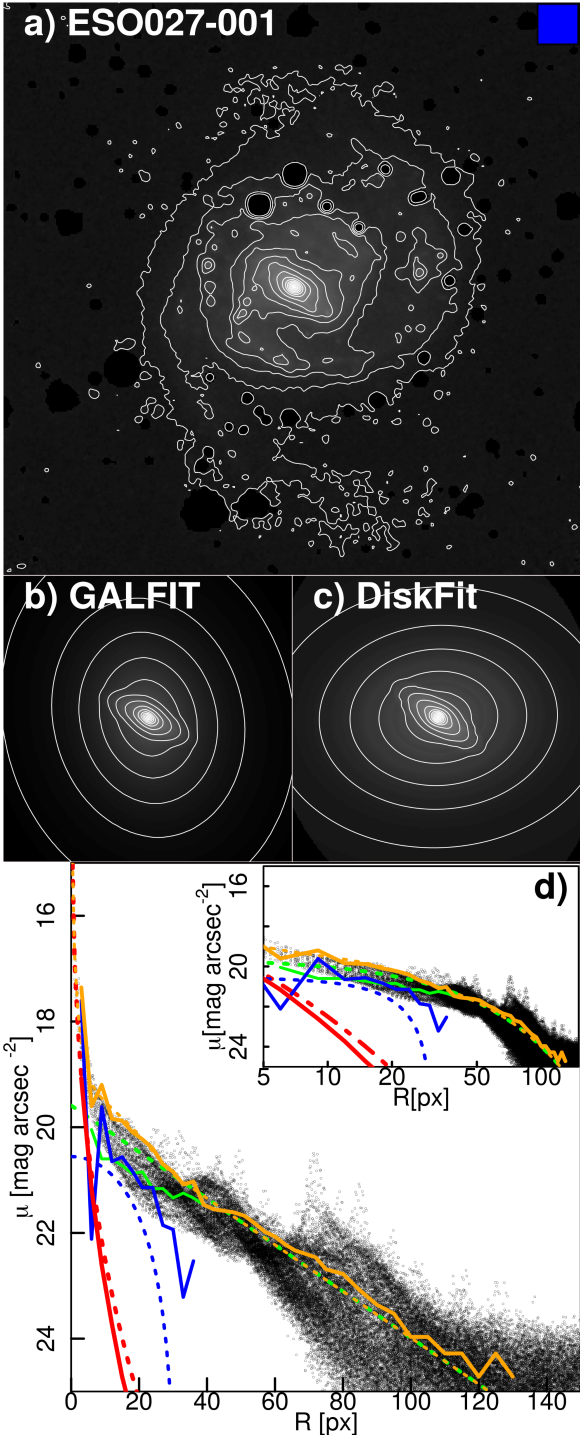}
	\caption{Same as in Figure \ref{Figure7}, but for ESO027-001. ESO027-001 is an example of a comparison sample galaxy with significantly different disks recovered between the two algorithms.}
   \label{Figure8}
\end{figure}
\begin{figure}
	\centering
	\includegraphics[width=0.9\columnwidth]{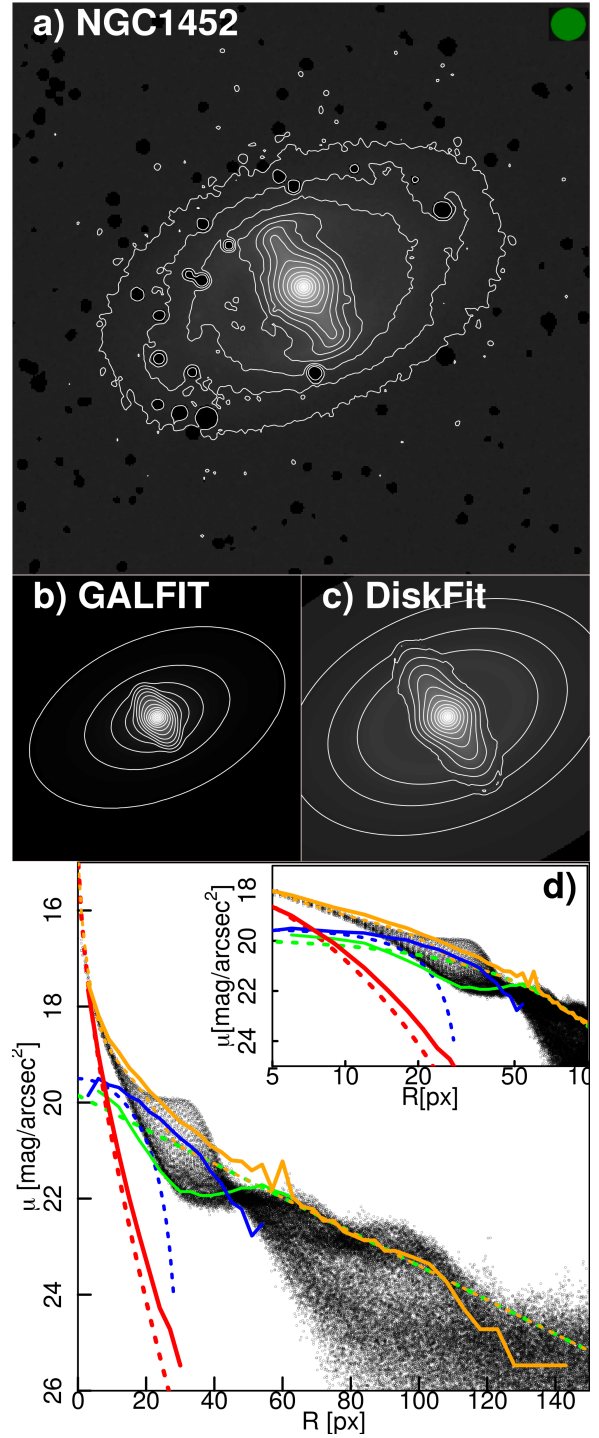}
	\caption{Same as Figure \ref{Figure7}, but for NGC1452. NGC1452 is an example of a comparison sample galaxy with significantly different bars recovered between the two algorithms. Note that the bar length was held fixed in the \textsc{galfit} model for this galaxy (\citetalias{2015ApJS..219....4S}).}
   \label{Figure9}
\end{figure}
\FloatBarrier
\begin{figure*}
	\includegraphics[width=\columnwidth]{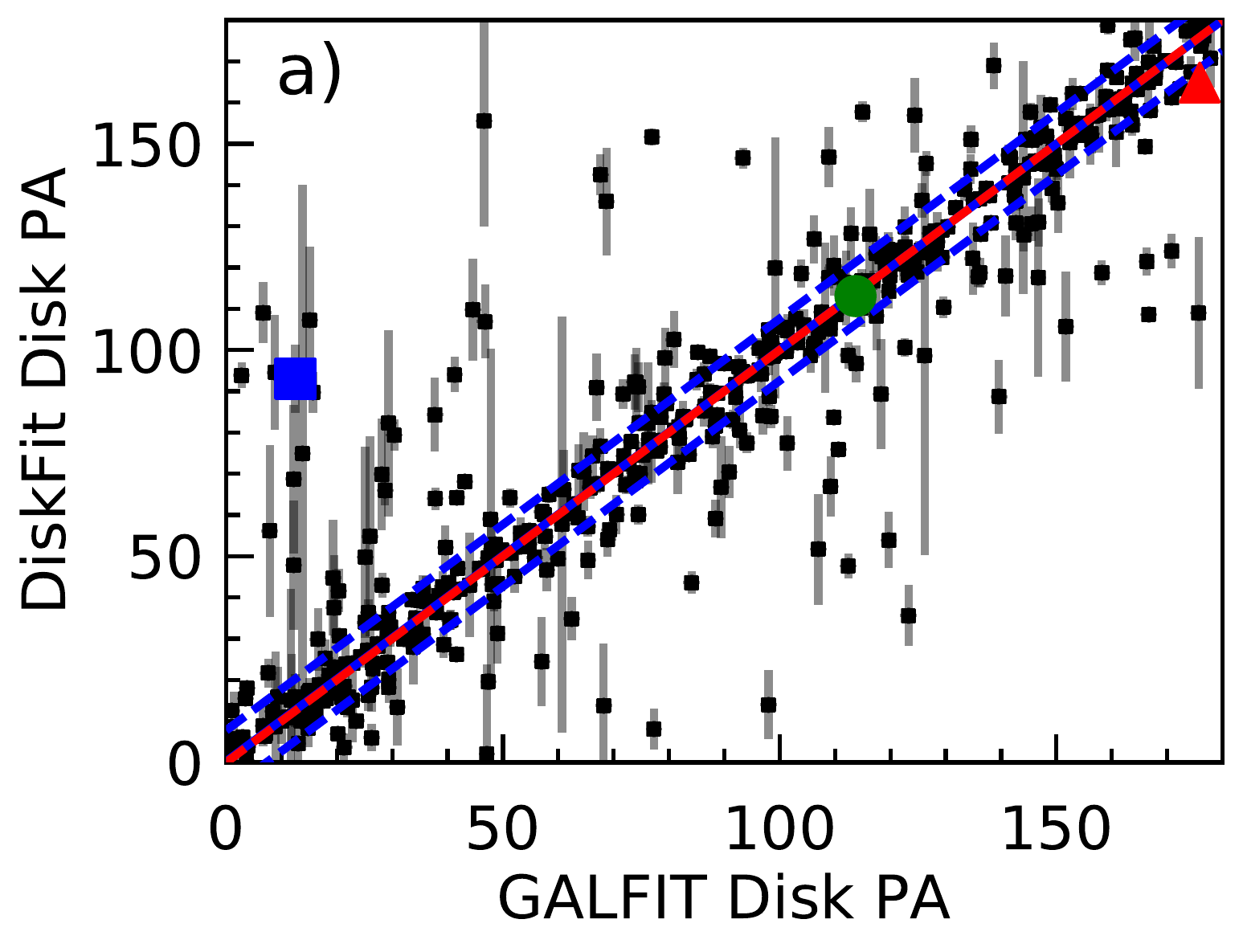}\includegraphics[width=\columnwidth]{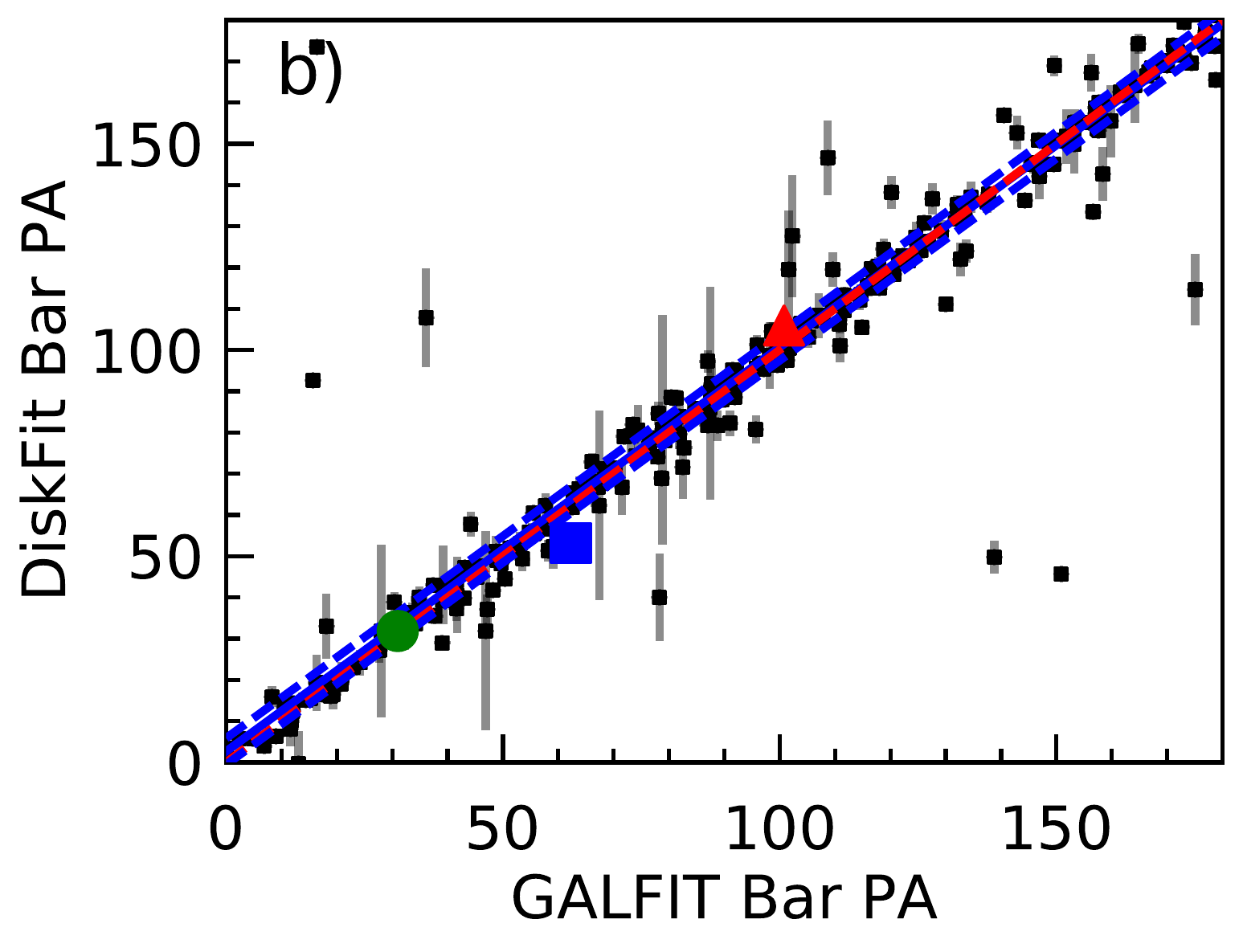}
	\includegraphics[width=\columnwidth]{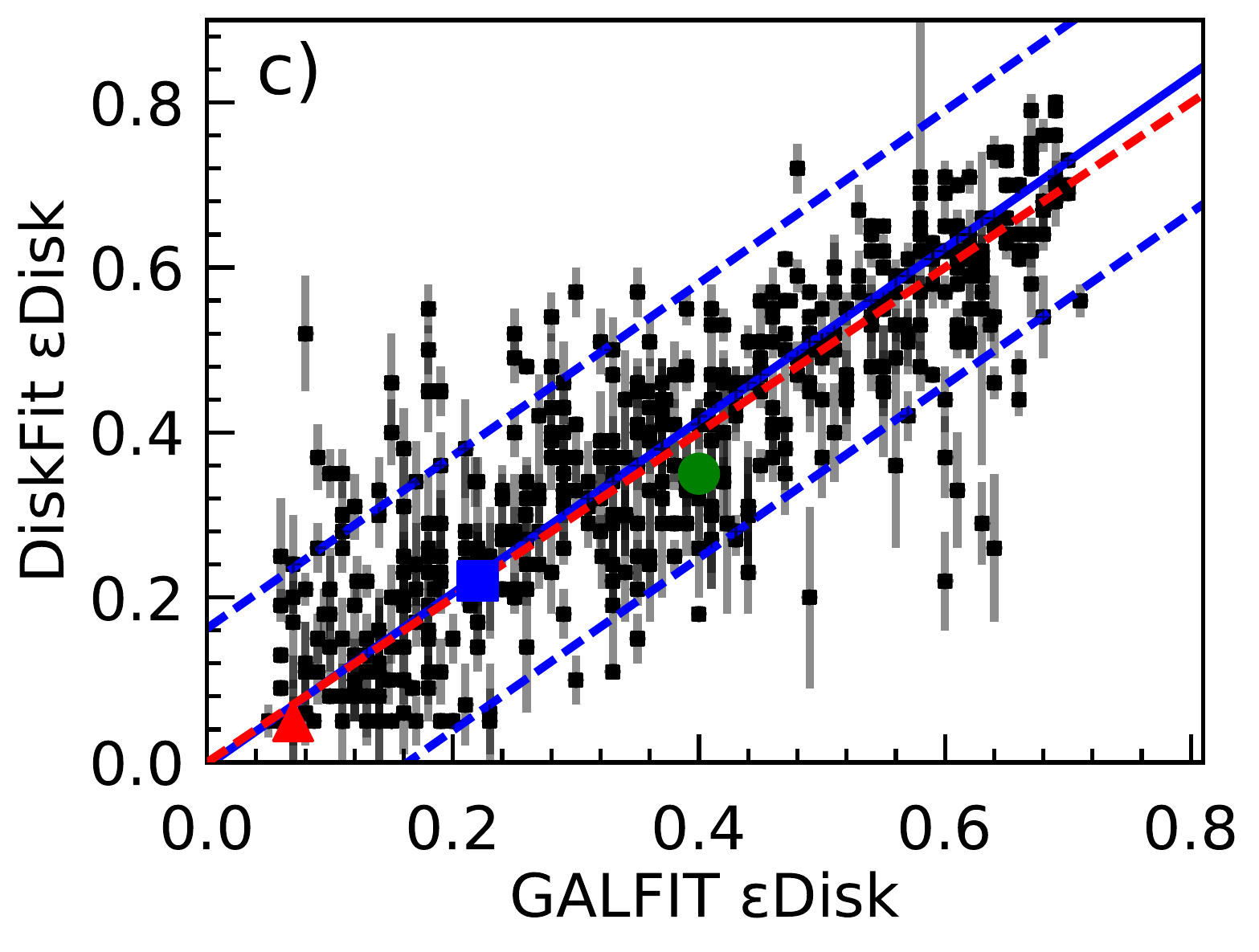}\includegraphics[width=\columnwidth]{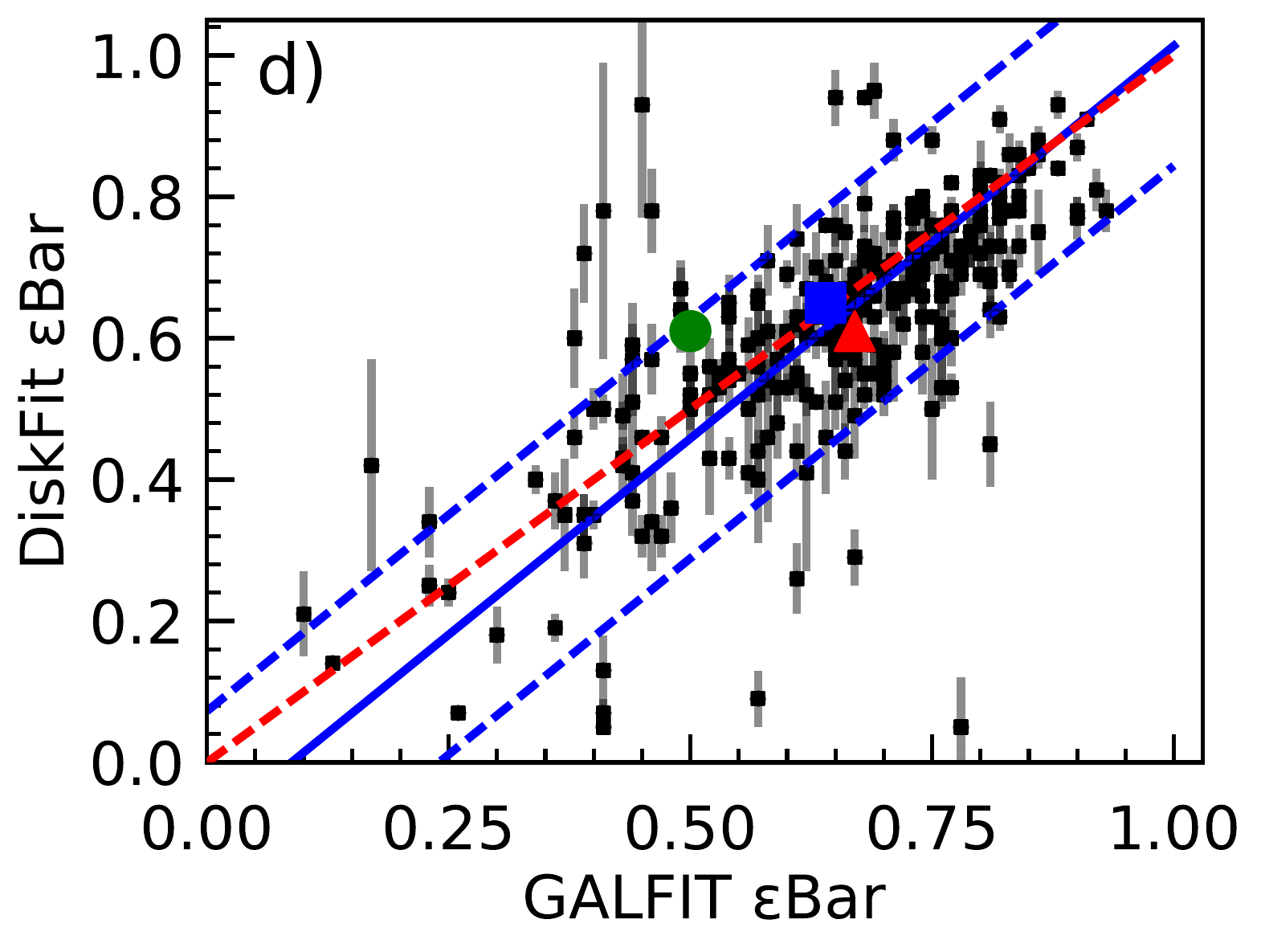}

	\caption{The position angle recovered by \textsc{DiskFit} for \textbf{a):} the disk as a function of the values adopted in the \textsc{galfit} models and \textbf{b):} the bar as a function of that recovered by \textsc{galfit}, and the ellipticity recovered by \textsc{DiskFit} for \textbf{c):} the disk and \textbf{d):} the bar as a function of that recovered by \textsc{galfit}. The dashed red line is the 1:1 relation, solid blue is the best-fitting linear least squares relationship, and the dashed blue lines show two times the normalized median absolute deviation (MAD/1.4826) about the best fit. The coloured points indicate the comparison sample galaxies shown in Figures \ref{Figure7} to \ref{Figure9}. \label{Figure10}}
\end{figure*}

\begin{figure}
	 %To include a figure from a file named example.*
	%or pdf, png, jpg if compiling using pdflatex
	\includegraphics[width=\columnwidth]{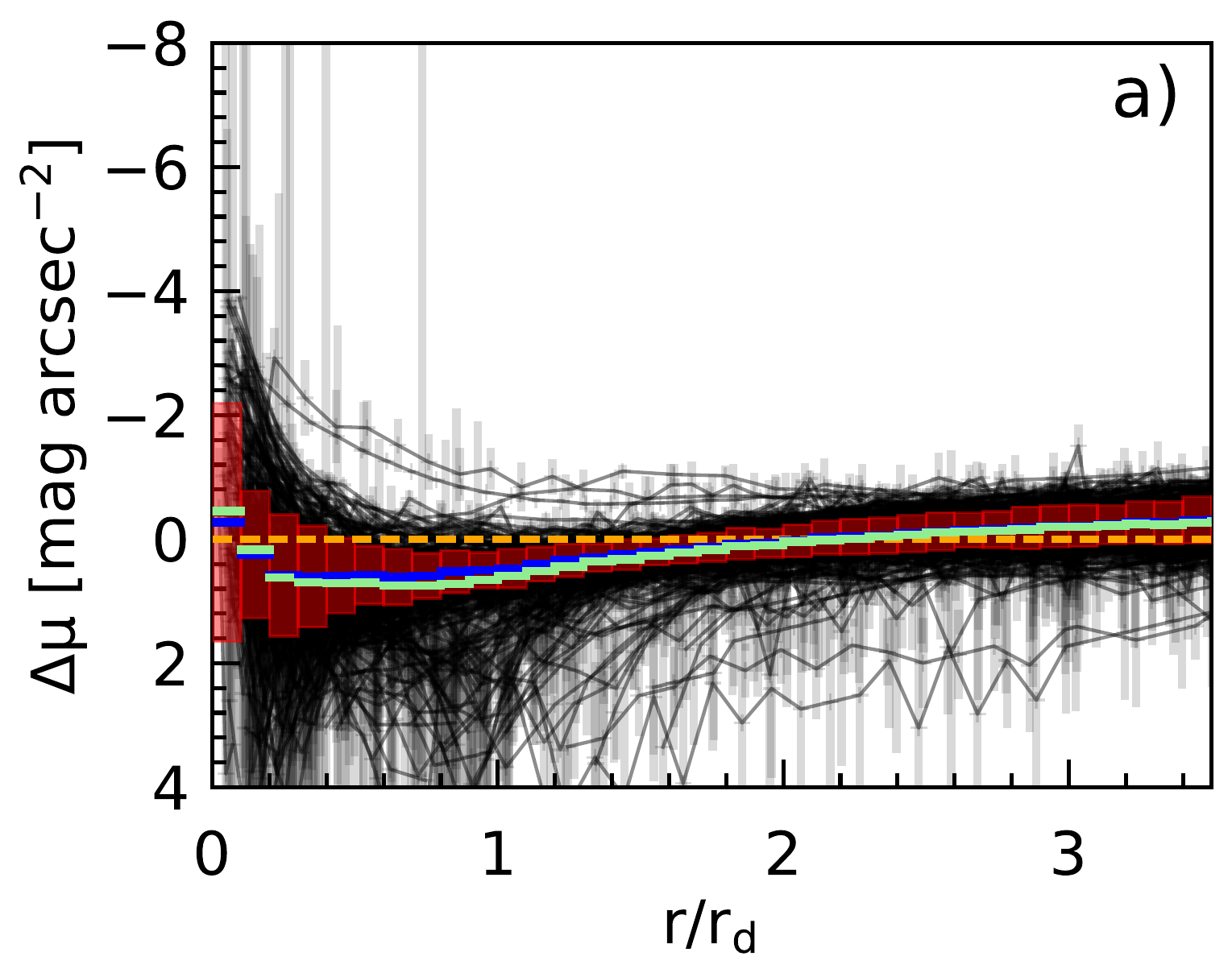}
	\includegraphics[width=\columnwidth]{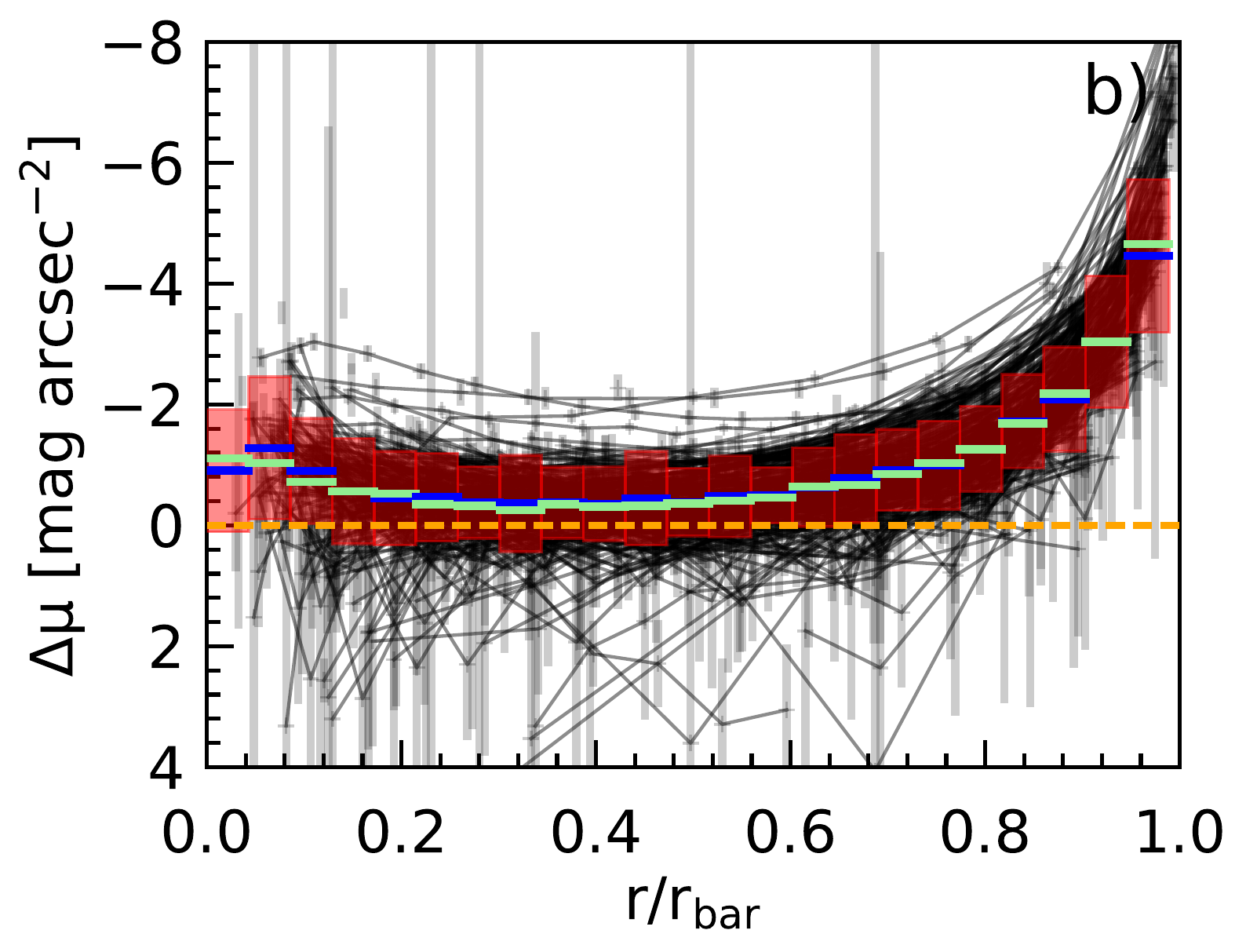}
	\caption{ The difference between the surface brightness measured by \textsc{DiskFit} and the \textsc{galfit} profiles for \textbf{a):} bar surface brightness profiles (Ferrers bars), and \textbf{b):} disk surface brightness profiles (exponential disks). The dashed yellow line indicates the point where the difference is zero. The red bars show the interquartile range, with the blue horizontal lines indicating the median within each bar and light green showing the mean within each bar. Note that the range of y-values in these figures are approximately twice that of Figure \ref{Figure6}, and that the values for $r_{bar}$ and $r_d$ are those recovered by \citetalias{2015ApJS..219....4S} using \textsc{galfit}. In this plot we exclude the galaxies fit by two disks in \citetalias{2015ApJS..219....4S}.\label{Figure11} }
\end{figure}
 \begin{figure}
	\centering
	\includegraphics[width=0.9\columnwidth]{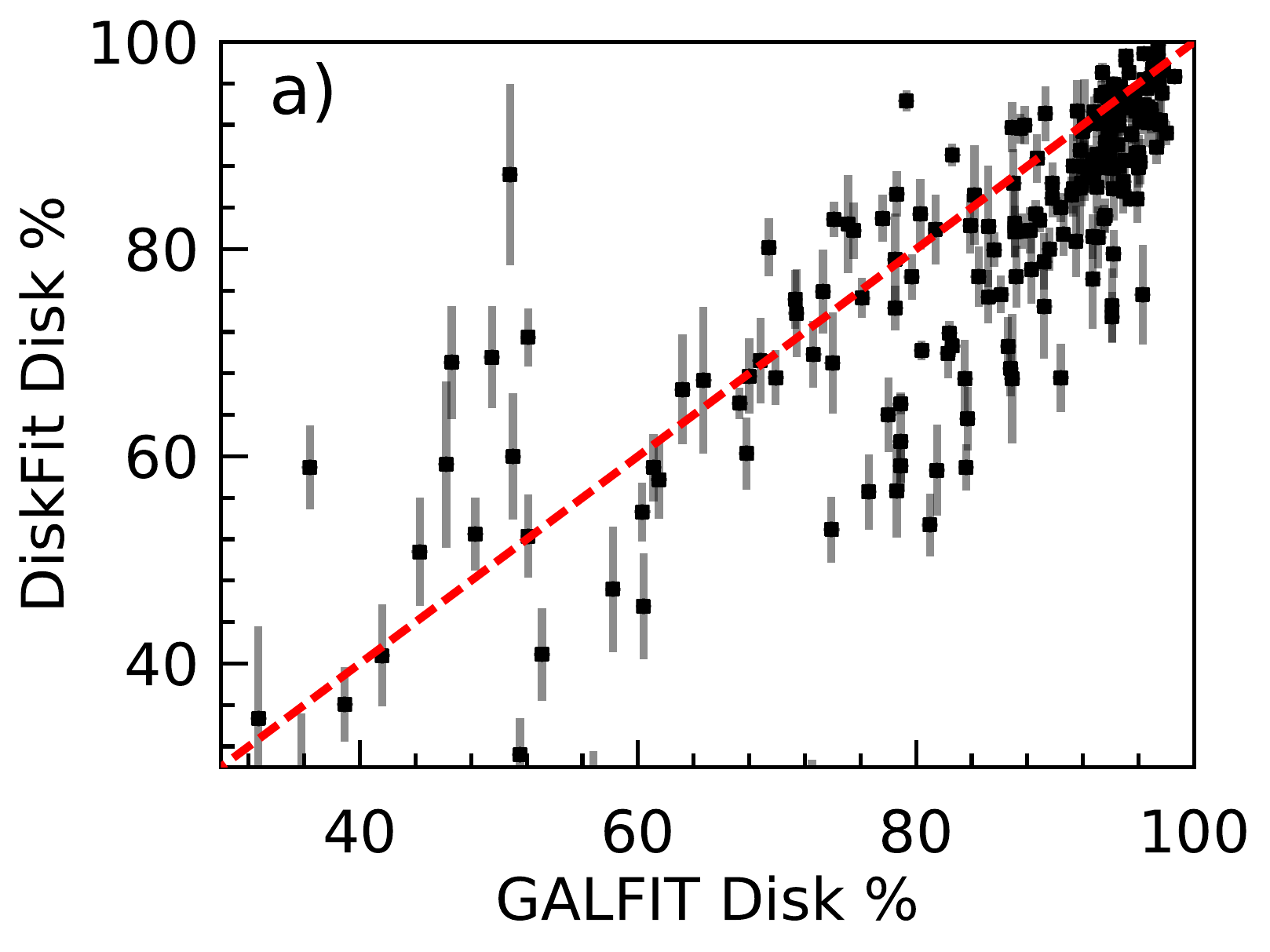}
	\includegraphics[width=0.9\columnwidth]{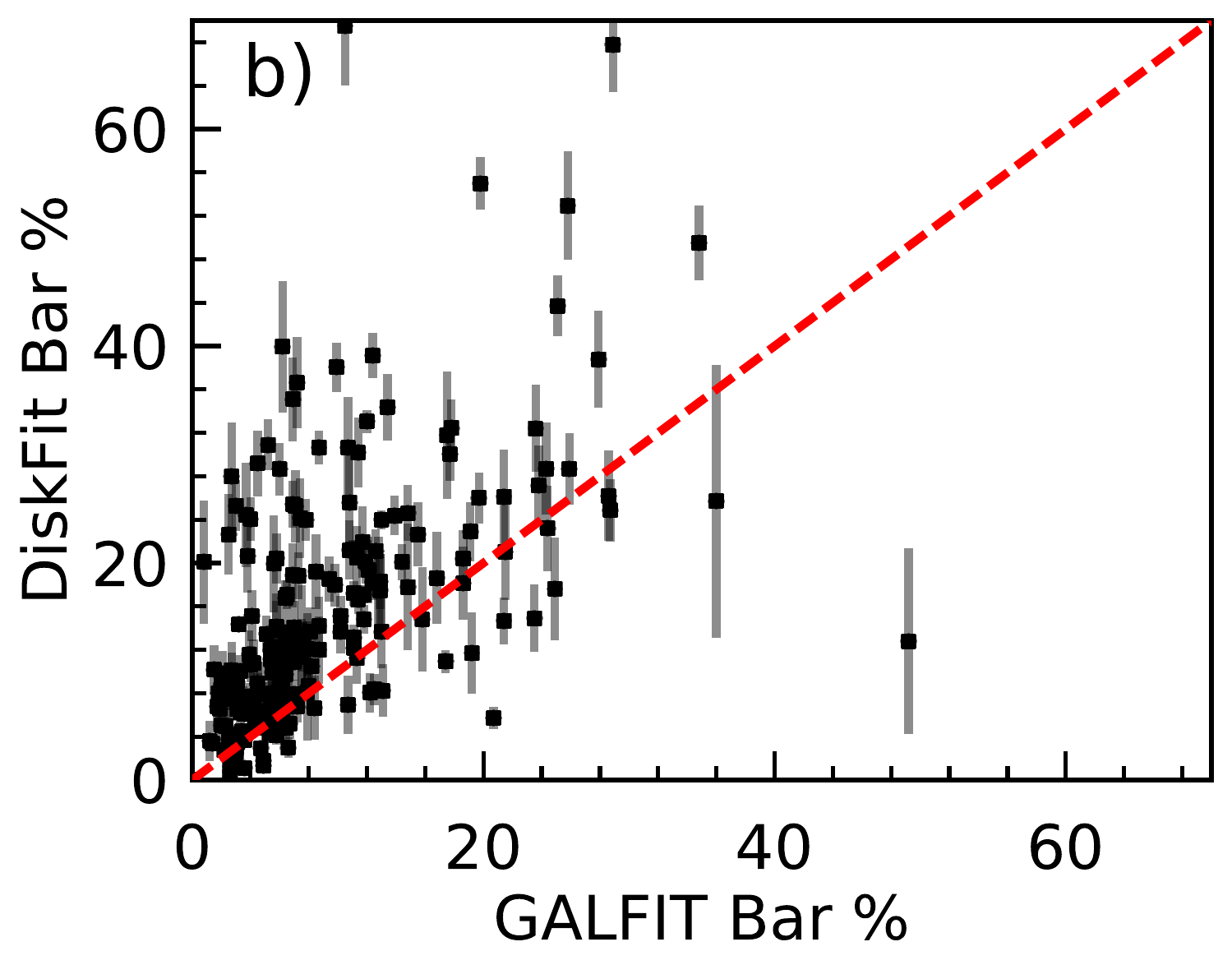}
	\caption{The fraction of light recovered within \textbf{a):} the disk and \textbf{b):} the bar by \textsc{DiskFit} as a function of that recovered by \textsc{galfit}. The dashed red line is 1:1 relation.  \label{Figure12}}
\end{figure}
\subsection{Fits to Real Galaxies: Comparing \textsc{DiskFit} and GALFIT}\label{section4_2}

This section focuses on comparing the best-fitting \textsc{DiskFit} models of the comparison sample to those adopted by \textsc{galfit} in \citetalias{2015ApJS..219....4S}. As explained in \S \ref{section4_1}, we include the same model components as in the \citetalias{2015ApJS..219....4S} decompositions and use their best fitting values as input guesses for \textsc{DiskFit}. We therefore expect the resulting \textsc{DiskFit} models to resemble the \textsc{galfit} ones as much as possible given the differences in modelling approach. 

Figures \ref{Figure7}-\ref{Figure9} show representative examples of fits to individual galaxies, while Figures \ref{Figure10} and \ref{Figure11} examine the results for the comparison sample as a whole. Since no uncertainties on the \citetalias{2015ApJS..219....4S} models are available, Figure \ref{Figure10} shows only vertical error bars while those in Figure \ref{Figure11} only account for \textsc{DiskFit} uncertainties.  For comparison sample galaxies modelled with two disks by \citetalias{2015ApJS..219....4S}, we plot the properties of one with the largest $r_{eff}$ in Figure \ref{Figure10}. We have verified that the properties of galaxies with nuclear components or second disks (flags `b' and `c' in Table \ref{table1}), exhibit quantitatively and qualitatively similar behaviour in Figures \ref{Figure10} and \ref{Figure11} as the other comparison sample galaxies. Parameters of the best-fitting linear relationships between our \textsc{DiskFit} fits and those adopted by \citetalias{2015ApJS..219....4S}'s models are shown in Table \ref{table_stats}.

Figure \ref{Figure7} shows an example of a galaxy, NGC1022, for which the best fitting \citetalias{2015ApJS..219....4S} \textsc{galfit} (panel b) and \textsc{DiskFit} (panel c) disk-bulge-bar models closely resemble each other. The surface brightness profiles of individual components (panel d) are also similar, with \textsc{DiskFit} trading some disk light (solid green line) for bar light (solid blue line) at 10px $<r<$ 30px (7.5 arcsec < $r$ < 23 arcsec) compared to \textsc{galfit} (dashed green and blue lines, respectively). NGC1022 is indicated by the red triangles in Figure \ref{Figure10}; the \textsc{DiskFit} and \textsc{galfit} disk and bar position angles and ellipticities are similar. We note that the disk ellipticities reported by \citetalias{2015ApJS..219....4S} are derived from the shape of the outer isophotes of the images, and not by \textsc{galfit} itself. Nonetheless, we find good agreement between the DiskFit models and the \citetalias{2015ApJS..219....4S} ones for galaxies without pronounced spiral arms, rings, or other features that are not included in the models.

Figures \ref{Figure8} and \ref{Figure9} illustrate that when such features are present in a galaxy, the best fitting \citetalias{2015ApJS..219....4S} \textsc{galfit} and \textsc{DiskFit} models can differ significantly. By construction, DiskFit uses the entire radial range of the disk to determine the PA \citep{2007ApJ...664..204S}. ESO027-001 in Figure \ref{Figure8} has pronounced spiral arms that pull on the fits differently resulting in disks with discrepant PAs as shown by the blue squares in Figure \ref{Figure10}. NGC1452 in Figure \ref{Figure9} exhibits a ring structure at the end of the bar that also produces different \textsc{galfit} and \textsc{DiskFit} models, with the latter returning a much longer, brighter bar than the former with a slightly higher $\epsilon Bar$ as shown by the green circles in Figure \ref{Figure10}. We find that the photometric decompositions of galaxies with spiral arms and rings are the ones most likely to differ significantly when modelled with \textsc{DiskFit} versus \textsc{galfit}.

Figure \ref{Figure10} shows that, on the whole, the two algorithms recover similar disk and bar geometries, with a much less centrally concentrated distribution of points than that seen in the synthetic galaxies (compare to the spread in Figure \ref{Figure5}c) and \ref{Figure5}d)). The trend lines are consistent with the 1:1 relations within the uncertainties (see Table \ref{table_stats}). The agreement between the position angles and ellipticities of the bars and disks recovered by \textsc{DiskFit} and adopted by \textsc{galfit} in \citetalias{2015ApJS..219....4S} is good for the majority of the sample galaxies. There are many more extreme outliers than that seen in the simulated sample (Figure \ref{Figure5}a) for the disk PA, resulting in a significantly increased MAD. Many of the points that lie outside of the normalized MAD for all quantities have small error bars, indicating significantly different best-fitting models for the two algorithms (see Figure \ref{Figure8}).
 
Figure \ref{Figure11} compares the difference between the bar and disk surface brightness profiles returned by \textsc{galfit} and \textsc{DiskFit}. Here, $\Delta \mu$ represents \textsc{DiskFit}'s best-fitting surface brightness profile minus the profile recovered by \citetalias{2015ApJS..219....4S} using \textsc{galfit}: $\Delta \mu=\mu_{\textsc{DiskFit}}-\mu_{\textsc{galfit}}$. Points with $\Delta\mu < 0$ correspond to regions where \textsc{DiskFit} attributes more light than \textsc{galfit}, and points with $\Delta\mu > 0$ correspond to regions where \textsc{DiskFit} attributes less light (note the units of mag arcsec$^{-2}$). We normalize the values radially by $r_{bar}$ and $r_{d}$ returned by \textsc{galfit}. For simplicity in this comparison, we exclude sample galaxies modelled by \citetalias{2015ApJS..219....4S} using two disks.

There is good agreement between the disks recovered by \textsc{DiskFit} and \textsc{galfit} for $r > 2r_d$ in Figure \ref{Figure11}a), although \textsc{DiskFit} attributes slightly more light to the disk as $r$ increases. For $r<1.5r_d$ in the sample as a whole, there is more light in the exponential \textsc{galfit} disks than their non-parametric \textsc{DiskFit} counterparts; this is also the case in the individual models shown in Figures \ref{Figure7}-\ref{Figure9}. Figure \ref{Figure11}b) shows that this light is being attributed the bar, which has a higher surface brightness for $0<r<r_{bar}$ in the \textsc{DiskFit} models compared to the \textsc{galfit} models. A larger $\Delta_{\mu}$ is found for $r > 0.8 r_{bar}$, where the non-parametric \textsc{DiskFit} bars are several times brighter than the Ferrers \textsc{galfit} bars (albeit in a region where the bar itself is faint). We note that this behaviour is not seen in the fits to simulated galaxies (c.f. Figure \ref{Figure6}), where \textsc{DiskFit} cleanly recovers true exponential disks and Ferrers bars. This suggests that the surface brightness distributions in the \sfg{}   galaxies are more complicated than the sum of these two components. 

We note that cumulative properties derived from photometric fits are even more uncertain. Figure \ref{Figure12} compares the light fractions implied by our \textsc{DiskFit} and the \citetalias{2015ApJS..219....4S} \textsc{galfit} fits for comparison sample galaxies that include both components: unless the fraction of light in the disk exceeds $\simeq 85\%$, the relative brightnesses of the disk and bar components can differ by over  $30\%$.

\begin{table}
\centering
\begin{tabular}{|c|c|c|c|}
\hline
        Parameter &  & &\\ 
	(Figure \ref{Figure5}) &  $\alpha$ & $\beta$ & MAD \\
        (1) & (2) & (3) & (4) \\
\hline
        $\epsilon$Disk  & 1.000(0.007)    & -0.003(0.003) & 0.0043  \\
        $\epsilon$Bar   & 0.965(0.008)    & 0.025(0.005)  & 0.014   \\
        Disk PA ($^o$)  & 1.0000(0.0001)  & -0.009(0.004)  & 0.25    \\
        Bar PA ($^o$)   & 1.0000(0.0001)  & -0.018(0.003)  & 0.36    \\
\hline
        Parameter &  & & \\
	(Figure \ref{Figure10}) & $\alpha$ & $\beta$ & MAD \\
\hline
	$\epsilon$Disk & 1.05(0.02)       & -0.01 (0.01) & 0.083 \\
	$\epsilon$Bar  & 1.11(0.03)       & -0.10 (0.02) & 0.085 \\
	Disk PA ($^o$) & 0.998(0.003)     & 0.2   (0.2)  & 7.5 \\
	Bar PA ($^o$)  & 0.98(0.01)       & 3     (1)    & 3.2 \\
\hline
\end{tabular}
	\caption{Best-fitting linear least squares regression parameters for relations shown in Figure \ref{Figure5} (synthetic sample) and in Figure \ref{Figure10} (\sfg{} subsample). Column 1: parameter of interest. Columns 2 and 3: best-fitting slope ($\alpha$) and intercept ($\beta$) from a linear least-squares fit $P_{DF} = \alpha P_x + \beta$, where $P_{DF}$ is the best-fitting \textsc{DiskFit} value and $P_x$ is the true value in the Figure \ref{Figure5} trends, and the value adopted by \textsc{galfit} in \citetalias{2015ApJS..219....4S} in the Figure \ref{Figure10} trends (numbers in parentheses are the uncertainties on the reported values). Column 4: median absolute deviation of the best-fitting \textsc{DiskFit} values. \label{table_stats}}
\end{table}
%Parameter &  (Average Error) & (Standard Deviation) & $\alpha$(Slope) & (Intercept) \\ 

%The difference between the \textsc{DiskFit} and \textsc{galfit} disks is not large, with \textsc{DiskFit} recovering a slightly increasing brighter disk at radii beyond $\sim 2 r_d$ than that recovered by \textsc{galfit}, and a slightly dimmer disk within $2r_d$. The most important feature to note is in $\Delta \mu$ of the bar. There is a widespread significant discrepancy between the two algorithms toward the end of the bar, starting as early as 0.75$r/r_{bar}$ and increasing outward. The dimmer trend in the disk recovered by \textsc{DiskFit} within $2r_d$ is consistent with the discrepancy at the outer end of the bar. This interaction is not seen for the simulated galaxies, suggesting that the profile of a bar is more complicated than that suggested by the Ferrers profile used by \citetalias{2015ApJS..219....4S} and in our simulations.

\section{Discussion and Conclusions}\label{section5}
We have presented non-parametric photometric models of over 500 disk galaxies from \sfg{}   using \textsc{DiskFit}. We use a suite of simulated galaxies embedded in \sfg{}-like images to validate \textsc{DiskFit}'s performance on our 570-galaxy \sfg{}   subsample (\S \ref{section3_2}, Figures \ref{Figure2} and \ref{Figure3}), and find that \textsc{DiskFit} non-parametrically recovers the geometry and surface brightness distributions of exponential disks and Ferrers bars as long as their position angles differ by more than 5$^\circ$ (Figures \ref{Figure5} and \ref{Figure6}). By contrast, \textsc{DiskFit} does not reliably recover the S\'ersic bulge properties of our simulated galaxies (Figure \ref{Figure4}) and we ignore the bulge region in our subsequent \sfg{}   fits.

We then carry out \textsc{DiskFit} decompositions of the 532 \sfg{}   subsample galaxies with $|$PAdisk-PAbar$| >5^\circ$ as determined by \citetalias{2015ApJS..219....4S} and adopted by the parametric \textsc{galfit} algorithm  (Table \ref{table1}), and compare the best-fitting parameters to the \citetalias{2015ApJS..219....4S} values. While on the whole the \textsc{DiskFit} and \textsc{galfit} models return similar results, we find that discrepancies between the best-fitting disk and bar geometries of some systems well exceed the uncertainties that we estimate using a robust bootstrap resampling technique. Comparing the disk and bar surface brightness profiles, we find that \textsc{DiskFit} attributes more light to the bar and less to the disk than in the \citetalias{2015ApJS..219....4S} \textsc{galfit} models across the extent of the bar (Figure \ref{Figure11}). This difference is particularly striking for $r\gtrsim 0.8r_{bar}$, where the non-parametric \textsc{DiskFit} bars are several times brighter than the Ferrers \textsc{galfit} bars. 
%Despite adopting the same model components and initial parameter values as found by \citetalias{2015ApJS..219....4S} using \textsc{galfit} in our comparison sample, we find that the scatter between our best-fitting disk and bar geometries and those reported by \citetalias{2015ApJS..219....4S} is larger than the average parameter uncertainties that we estimate using a robust bootstrap-resampling technique as well as that found in our simulated galaxy fits (Figure \ref{Figure10}, Table \ref{table_stats}). 

The failure of \textsc{DiskFit} to recover bulges in simulated \sfg{}-like images highlights the importance of validating decomposition algorithms, especially before applying them to large samples of real systems \citep[e.g.][]{1987AJ.....93...60S,1995ApJ...448..563B,1999AJ....117.1219W,2003ApJ...582..689M,2008MNRAS.384..420G,2017MNRAS.469.3541P}. It is possible that the non-parametric nature of \textsc{DiskFit} makes bulge recovery more difficult than for parametric algorithms such as \textsc{galfit}, though \citet{2008MNRAS.384..420G}  finds that bulges with $r_{eff}$ comparable to the seeing radius are difficult to recover even when both the bulge and the disk are parametrized. We caution against the adoption of bulge parameters derived from 2D decompositions that do not include quantitative estimates of the associated uncertainties and algorithms limitations through simulations (e.g. \citetalias{2015ApJS..219....4S}, \citealt{1999AJ....117.1219W,2014ApJ...786..105J,2017MNRAS.471.1070B}).

The relatively large scatter between the best-fitting \textsc{DiskFit} parameters and those adopted in the \citetalias{2015ApJS..219....4S} models in our \sfg{}   comparison subsample suggests that, unless the (unreported) uncertainties in the \citetalias{2015ApJS..219....4S} \textsc{galfit} fits are considerably larger than those that we derive using \textsc{DiskFit} for the majority of the sample, different photometric decomposition approaches can return significantly different structural properties for disks and bars in nearby galaxies \citep[e.g.][]{1995ApJ...448..563B,2010AJ....139.2097P}. For \sfg{}   disk galaxies in particular, our \textsc{DiskFit} and \textsc{galfit} comparisons suggest that their disk and bar ellipticities are typically constrained to no better than $\Delta \epsilon \sim 0.1$ on average, while much larger uncertainties are implied by the outliers in Figure \ref{Figure10}. 

Our \textsc{DiskFit} models of the \sfg{}   comparison sample were constrained to have the same number of components and initial conditions as those reported by \citetalias{2015ApJS..219....4S} using \textsc{galfit}. While a detailed investigation of the impact of structural component and initial guess selection on model uncertainties is beyond the scope of this paper, we speculate that these effects would further widen the gap between parameters derived using different photometric modelling techniques. As explained in \S\ref{section4_1}, there are some sample galaxies for which \textsc{DiskFit} failed to converge on even a disk-only model when initiated using the best fitting values from \citetalias{2015ApJS..219....4S} as inputs. We suspect that this stems from the susceptibility of the Levenberg-Marquardt minimization scheme adopted by \textsc{DiskFit} and \textsc{galfit} to local minima in the  $\chi^2$ landscape. Indeed, recent kinematic model tests by \citet{2016MNRAS.455..754B} show that the optimization method itself may also affect the best-fitting values obtained from the same modelling algorithm. We therefore conclude that the differences we find between the \textsc{DiskFit} and \textsc{galfit} fits to the comparison sample galaxies likely underestimate the true uncertainties in \sfg{}   disk galaxy structural parameters recovered from 2D decomposition algorithms. 

In light of our simulations demonstrating that \textsc{DiskFit} accurately disentangles exponential disks from Ferrers bars, the differences between the surface brightness profiles recovered by \textsc{DiskFit} and the \citetalias{2015ApJS..219....4S} \textsc{galfit} models implies that the surface brightness distributions of the \sfg{}   comparison sample are not well-represented by these functional forms. The deficit of disk light that we find in the bar region is reminiscent of the  $'\Theta'$ shaped features recovered in some decompositions of real galaxies \citep[e.g.][]{2005MNRAS.362.1319L,2008MNRAS.384..420G,2016ApJS..225....6K}, as well as in simulations \citep[e.g.][]{2002MNRAS.330...35A}. At the same time, the excess of bar light recovered by \textsc{DiskFit} relative to the  \citetalias{2015ApJS..219....4S} \textsc{galfit} fits, particularly for $r>0.8r_{bar}$, implies that we non-parametrically recover brighter, longer bars than found in comparable Ferrers fits. We therefore concur with \citet{2017MNRAS.469.4414W} that real galactic bars may exhibit important differences from the Ferrers function form. This is particularly relevant when a bar's length is used as a proxy for its evolutionary state \citep{2011MNRAS.415.3627H,2015A&A...576A.102A,2016MNRAS.463.1074C}. It is possible that bar lengths derived from non-parametric 2D models such as \textsc{DiskFit} better represent their evolutionary states than do their Ferrers $r_{bar}$, but a careful validation of this hypothesis using mock and simulated galaxies is required \citep[e.g.][]{2002MNRAS.330...35A,2009A&A...495..491A}. We defer such an exploration using \textsc{DiskFit} to future work.

\section*{Acknowledgements}
Computations were performed on resources and with support provided by the Centre for Advanced Computing (CAC) at Queen's University in Kingston, Ontario. The CAC is funded by: the Canada Foundation for Innovation, the Government of Ontario, and Queen's University.

\noindent
KS acknowledges support from the Natural Sciences and Engineering Research Council of Canada (NSERC).
%The Acknowledgements section is not numbered. Here you can thank helpful
%colleagues, acknowledge funding agencies, telescopes and facilities used etc.

%%%%%%%%%%%%%%%%%%%%%%%%%%%%%%%%%%%%%%%%%%%%%%%%%%



\bibliographystyle{mnras}
\bibliography{main} % if your bibtex file is called example.bib

%\begin{thebibliography}{99}
%Salo H.., 2013, Journal of Improbable Astronomy, 1, 1
%\end{thebibliography}

%%%%%%%%%%%%%%%%%%%%%%%%%%%%%%%%%%%%%%%%%%%%%%%%%%

%%%%%%%%%%%%%%%%% APPENDICES %%%%%%%%%%%%%%%%%%%%%

\onecolumn
\appendix
\section{Full Table 1}		%uncommented in digital version
\begin{longtable}{ccccccccc}
\hline
Name        	& Model	& Disk PA [$^\circ$]	& $\epsilon$Disk	& Bar PA [$^\circ$]	& $\epsilon$Bar	& Disk \%	 & Bar \%	& flags\\
(1) & (2) & (3) & (4) & (5) & (6) & (7) & (8) & (9) \\
\hline
\endhead
ESO012-010  	& dbar	& $151(1)$ &  $0.62(0.02)$ &  $24(3)$ &  $0.44(0.04)$ &  $88(1)$ &  $12(1)$ &  \\
ESO013-016  	& dbar	& N/A 	& N/A 	& N/A 	& N/A 	& N/A 	& N/A 	& a \\
ESO026-001  	& dbar	& $45(14)$ &  $0.25(0.07)$ &  $71(1)$ &  $0.77(0.02)$ &  $92(1)$ &  $7.6(1.3)$ &  b \\
ESO027-001  	& bdbar	& $93(8)$ &  $0.22(0.06)$ &  $53(2)$ &  $0.65(0.05)$ &  $82(3)$ &  $12(3)$ &  \\
ESO027-008  	& d	& $145(2)$ &  $0.64(0.02)$ &  N/A& N/A& N/A& N/A& b \\
ESO048-017  	& dbar	& N/A 	& N/A 	& N/A 	& N/A 	& N/A 	& N/A 	& a b \\
ESO054-021  	& dbar	& $82(1)$ &  $0.50(0.03)$ &  $41(9)$ &  $0.37(0.04)$ &  $93(2)$ &  $7.4(2.0)$ &  \\
ESO079-005  	& dbar	& $183(2)$ &  $0.53(0.03)$ &  $13(1)$ &  $0.76(0.02)$ &  $74(4)$ &  $26(4)$ &  \\
\end{longtable}
% 489 38 144 89 19 16 8 686 116 570

      	%uncommented in digital version 
%\input{test.txt}		%uncommented in digital version
%\twocolumn			%uncommented in digital version

%%%%%%%%%%%%%%%%%%%%%%%%%%%%%%%%%%%%%%%%%%%%%%%%%%

% Don't change these lines
\bsp	% typesetting comment
\label{lastpage}
\end{document}